\documentclass[twocolumn,floats,groupedaddress,prb,aps]{revtex4}
\usepackage[T1]{fontenc}
\usepackage{lmodern}

\usepackage[english]{babel}

\usepackage{amsfonts}
\usepackage{amssymb}
\usepackage{graphicx}
\usepackage{dcolumn} 
\usepackage{amsmath}
\usepackage{bm}      
\usepackage{graphicx}
\usepackage{color}
\usepackage{epsfig,psfrag,subfigure,subeqnarray}
\usepackage{relsize}

\def\lan{\langle}
\def\ran{\rangle}
\def\va{\varepsilon}
\def\dag{\dagger}

\def\vk{{\bf k}}
\def\vR{{\bf R}}

\def\vp{{\bf p}}
\def\vq{{\bf q}}
\def\vr{{\bf r}}

\def\v0{{\bf 0}}

\def\vK{{\bf K}}

\newcommand{\bd}{\begin{equation}}
\newcommand{\ed}{\end{equation}}
\newcommand{\be}{\begin{equation}}
\newcommand{\ee}{\end{equation}}
\newcommand{\bt}{\begin{split}}
\newcommand{\et}{\end{split}}

\newcommand{\bn}{\begin{align}}
\newcommand{\en}{\end{align}}
\newcommand{\bea}{\begin{eqnarray}}
\newcommand{\eea}{\end{eqnarray}}
\newcommand{\ba}{\begin{array}}
\newcommand{\ea}{\end{array}}
\newcommand{\nn}{\nonumber}

\DeclareMathAlphabet\mathbfcal{OMS}{cmsy}{b}{n}

\begin{document}


\title{Ab initio quantum approach to ``electron-hole exchange'' 

for semiconductors hosting Wannier excitons}

\author{Monique Combescot}
\affiliation{Institut des NanoSciences de Paris, Sorbonne Universit\'e, CNRS, 4 place Jussieu, 75005 Paris}
\author{Thierry Amand}
\affiliation{Universit\'e de Toulouse, INSA-CNRS-UPS, 31077, Toulouse, France}
\author{Shiue-Yuan Shiau}
\affiliation{Physics Division, National Center for Theoretical Sciences, Taipei, 10617, Taiwan}

\begin{abstract}
We propose a quantum approach to ``electron-hole  exchange'', better named electron-hole \textit{pair} exchange, that makes use of the second quantization formalism to describe the problem in terms of Bloch-state electron \textit{operators}. This approach renders transparent the fact that such singular effect comes from interband Coulomb processes. We first show that due to the sign change when turning from valence-electron destruction operator to hole creation operator, the interband Coulomb interaction only acts on spin-singlet electron-hole pairs, just like the interband electron-photon interaction, thereby making these spin-singlet pairs optically bright. We then show that when written in terms of reciprocal lattice vectors $\textbf{G}_m$, the singularity of the interband Coulomb scattering in the small wave-vector transfer limit, entirely comes from the $\textbf{G}_m=\v0$ term, which renders its singular behavior easy to calculate. Comparison with the usual real-space formulation in which the singularity appears through a sum of ``long-range processes'' over all $\vR_\ell\neq\v0$ lattice vectors, once more proves that periodic systems are easier to handle in terms of reciprocal vectors $\textbf{G}_m$ than in terms of lattice vectors $\vR_\ell$. Well-accepted consequences of the ``electron-hole exchange'' on excitons and polaritons are reconsidered and refuted for different major reasons.
\end{abstract}
\date{\today}
\maketitle

\section{Introduction}
Excitons result from semiconductor excitations that are correlated by Coulomb interaction. In materials hosting Wannier excitons, the electron energies form bands separated by gaps, as beautifully demonstrated by the Bloch theorem for periodic crystals\cite{Kittelbook,Bloch1929,Merminbook}. The physically relevant bands are the highest valence band and the lowest conduction band, respectively full and empty for undoped semiconductors at zero temperature. The first set of semiconductor excitations corresponds to one valence electron  jumping to the conduction band, with an empty state left in the valence band. It is possible to show that a full valence band minus one electron essentially behaves as a single particle called ``hole'', that  has a positive charge and a positive mass. Yet, the remaining valence electrons can ``boil'' into virtual conduction electron-valence hole pairs. Such virtual pairs are the ones that lead to reduce the Coulomb potential through a dielectric constant\cite{Monicbook} of the order of 10.

The Coulomb interaction acts on valence and conduction electrons in two distinct ways\cite{Monicbook}.

\noindent(1) Each electron can stay in its band through \textit{intraband} Coulomb processes (see Fig.~\ref{fig_5}). Their repetition transforms one plane-wave conduction electron and one plane-wave valence hole into a single plane wave for the exciton center of mass.

\noindent(2) Each electron can change band through \textit{interband} Coulomb processes (see Fig.~\ref{fig_6}): A conduction electron returns to an empty state of the valence band, while a valence electron jumps to the conduction band, leaving an empty state in the valence band. Since the valence band with an empty state is nothing but a hole, the interband processes  correspond to the recombination of an electron-hole pair along with the excitation of another pair (see Fig.~\ref{fig_8}). So, the interband Coulomb interaction fundamentally leads to an exchange of electron-hole \textit{pairs}. This literally differs from an ``electron-hole exchange'', as commonly named, because an electron cannot have a quantum exchange with a different fermion like the hole. Actually, an even better name simply is ``interband Coulomb interaction'' because this name readily tells the physics that drives the effect.

The interband Coulomb processes have a marginal role compared with the intraband processes that correlate free electron-hole pairs into a Wannier exciton. Yet, interesting effects induced by these interband processes deserve investigation.
 
\noindent(\textit{i}) The interband Coulomb interaction only acts on electron-hole pairs that are in a spin-singlet state, just like the interband electron-photon interaction, thereby making these pairs optically bright, whereas pairs in a spin-triplet state are dark, that is, not coupled to photons.

\noindent(\textit{ii}) When the valence band has a threefold spatial level, as for GaAs-like semiconductors, the interband Coulomb scattering is highly singular when the wave-vector transfer --- which also is the wave vector of the scattered electron-hole pair --- goes to zero: It has  two different limits that depend on the direction of this wave vector with respect to the crystal axes.

The singularity of the interband Coulomb scattering is commonly understood in terms of ``long-range'' and ``short-range''  processes  within the usual lattice vector formulation: Long-range processes that take place across multiple lattice cells are responsible for the singularity, while short-range processes that  take place inside a single lattice cell brings a regular contribution. Although adopted for  a long time, we will show that this formulation is not the appropriate one to pin down the scattering singularity for the very simple reason that the lattice vector space is not the appropriate space to handle periodic systems.

\textbf{In this work, we present} an \textit{ab initio} approach to the various scatterings associated with the Coulomb interaction in a semiconductor; it makes use of the second quantization formalism in terms of operators for Bloch-state electrons. Beside avoiding heavy Slater determinants, this operator formalism allows us to trivially elucidate why electron-hole pairs in a spin-\textit{triplet} state do not suffer the \textit{interband} Coulomb interaction. Moreover, through this Bloch-state formulation, it becomes easy to catch why the interband Coulomb scattering between electron-hole pairs has a singularity that depends on the direction of the pair center-of-mass wave vector with respect to the crystal axes.

The paper is organized as follows:

\noindent $\bullet$ In Sec.~\ref{sec1}, we come back to the foundation of the two-body Coulomb interaction between fermions in the conduction and valence bands, first in terms of valence and conduction electrons, and then in terms of electrons and holes, the latter being the appropriate language when dealing with excitations. We visualize this interaction through Feynman diagrams, which are especially enlightening in the case of interband Coulomb processes because they render transparent the fact that such processes fundamentally correspond to exchange an electron-hole \textit{pair}.     

\noindent $\bullet$ In Sec.~\ref{sec2}, we formulate the electron-electron Coulomb interaction in second quantization within the Bloch-state basis, which is the relevant one-electron basis for semiconductors hosting Wannier excitons. We pay a particular attention to the interband Coulomb scattering. First formulated in terms of lattice vectors $\vR_\ell$, we show how to rewrite this scattering in terms of reciprocal vectors $\textbf{G}_m$, which is the appropriate space to handle the lattice periodicity. 
We also pin down the importance of the one-body average electron-electron interaction, introduced to properly define the Bloch-state basis, as it eliminates the zero-wave-vector transfers not only for intraband Coulomb processes but also for interband processes --- a crucial point to get rid of spurious volume-infinite terms that appear in the calculation. To the best of our knowledge, this elimination has never been carefully established. Finally, we transform electron-electron interaction into electron-hole interaction. This change readily reveals that electron-hole pairs that suffer interband processes are in a spin-singlet state.

\noindent $\bullet$ In Sec.~\ref{sec3}, we analytically calculate the interband Coulomb scattering. We derive its singular behavior in the limit of small wave-vector transfer, this wave vector transfer also being  the center-of-mass wave vector of the scattered electron-hole pair. By using the expression of the interband scattering as a sum over reciprocal vectors $\textbf{G}_m$, we show that its singularity only comes from the $\textbf{G}_0=\v0$ term. To make link with the former approach to electron-hole exchange, we also calculate this scattering as a sum over lattice vectors $\vR_\ell$. The singularity then comes from the sum over all nonzero lattice vectors $\vR_\ell\neq\v0$, known as ``long-range processes''. We prove that this singularity is the same as the one coming from the  $\textbf{G}_0$ term of the $\textbf{G}_m$ sum. Bridging calculations done in the $\vR_\ell$ and $\textbf{G}_m$ spaces provides a deeper insight to this highly singular effect.

\noindent $\bullet$ In Sec.~\ref{sec3}, we present a ``state-of-the-art'' on the so-called ``electron-hole exchange''. We also reconsider the well-accepted consequences of this singular interband scattering in two fully different frameworks: The energy splitting between dark and bright excitons and the transverse-longitudinal splitting of the polariton. We explain why in both cases, but for different reasons, the singular electron-hole pair exchange can hardly be associated with experimental results.

\noindent $\bullet$ We then conclude.

\begin{figure}[t!]
\centering
\includegraphics[trim=0.5cm 8.5cm 0cm 2cm,clip,width=3.6in]{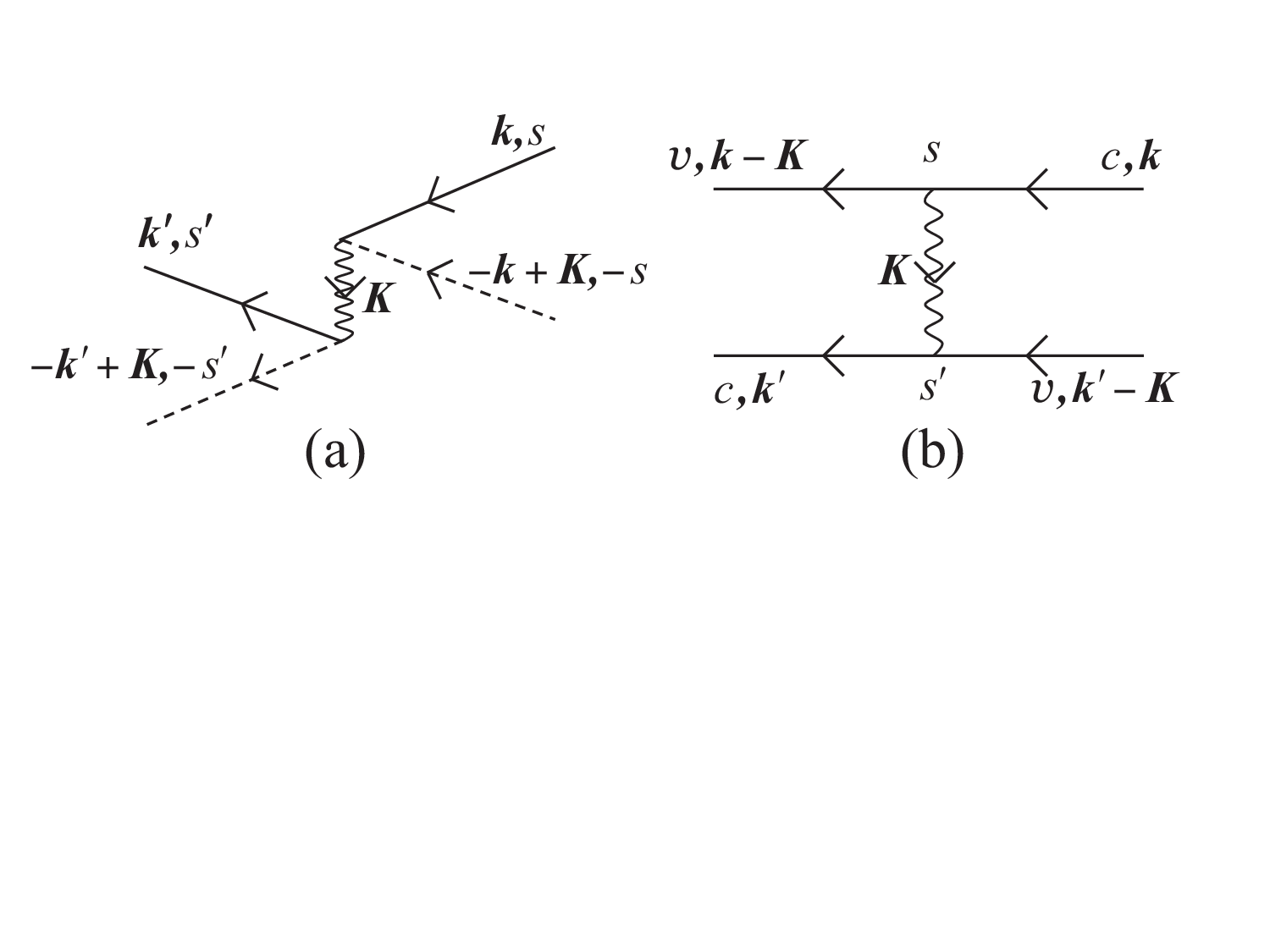}
\vspace{-0.7cm}
\caption{Interband Coulomb scattering $V_\vK(\vk',\vk)$ in terms of (a) electron (solid line) and hole (dashed line) pairs with center-of-mass wave vector $\vK$, or (b) valence and conduction electrons. Note that the wave-vector transfer $\vK$ for interband Coulomb processes also is the center-of-mass wave vector of the scattered electron-hole pair.  As Coulomb interaction conserves the spin, the electron-hole pairs involved in the interband Coulomb processes have a total spin equal to zero. They moreover are in a spin-singlet state due to a subtle sign change (see Eq.~(\ref{cheh_ex:14_1})) that appears when turning from valence electron to hole. }
\label{fig_1}
\end{figure}
\begin{figure}[t!]
\centering
\includegraphics[trim=0cm 11.4cm 11.4cm 2cm,clip,width=3in]{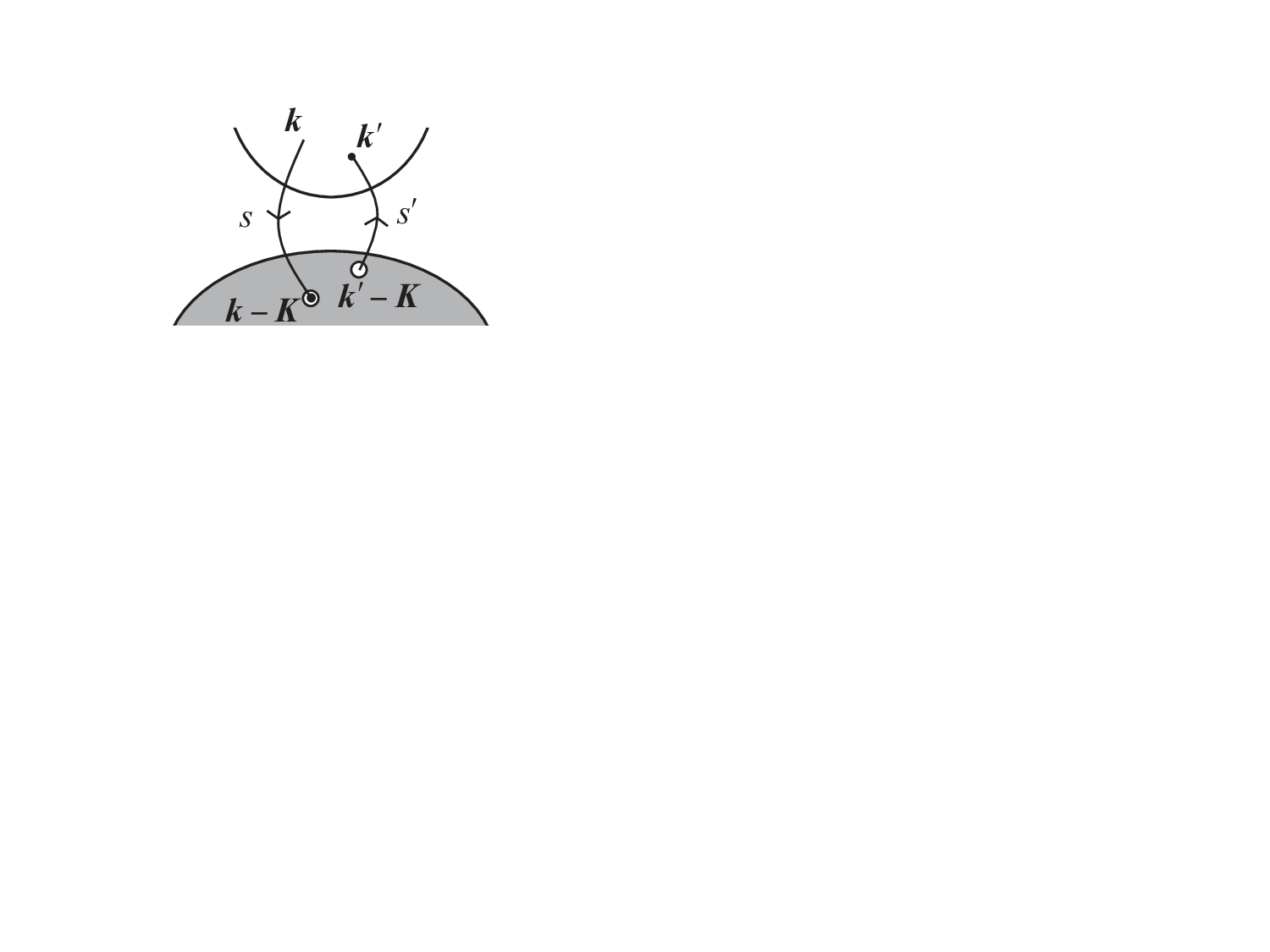}
\vspace{-0.7cm}
\caption{In the interband Coulomb process shown in Fig.~\ref{fig_1}(b), a spin $s$ electron goes from the conduction state $\vk$ to the valence state $(\vk-\vK)$, thus filling a $-(\vk-\vK)$ hole with spin $-s$ present in the valence band, while another electron having a $s'$ spin goes from the valence state $(\vk'-\vK)$ to the conduction state $\vk'$, leaving a $-(\vk'-\vK)$ hole with spin $-s'$ in the valence band. }
\label{fig_2}
\end{figure}

\section{Coulomb interaction between conduction and valence electrons\label{sec1}}

\subsection{Definitions}

What is commonly called ``electron-hole exchange'' refers to a two-body \textit{interband} process mediated by the Coulomb interaction. The system starts with an electron-hole pair having a center-of-mass wave vector $\vK$ and a $\vk$ electron; it ends with a pair having the same center-of-mass wave vector $\vK$ and a $\vk'$ electron. The state change involves filling a valence  hole with the $\vk$ conduction electron, while creating another hole and another $\vk'$ conduction electron, the initial and final electron-hole pairs having the same center-of-mass wave vector $\vK$ because the Coulomb interaction conserves the total wave vector of the involved pair. Let us denote as
\be\label{1}
V_\vK(\vk',\vk)
\ee
the scattering amplitude associated with this state change. It depends on three wave vectors: The pair center-of-mass wave vector $\vK$ and the wave vectors $\vk$ and $\vk'$ of the incoming and outgoing electrons.

 This interband process can be represented either in terms of electron and hole by the Feynman diagram of Fig.~\ref{fig_1}(a), or in terms of conduction and valence electrons by the Feynman diagram of Fig.~\ref{fig_1}(b). The corresponding physical process is shown in Fig.~\ref{fig_2}.

The purpose of this work is to derive the  behavior of $V_\vK(\vk',\vk)$ in the $\vK\rightarrow \bf0$ limit.

\begin{figure}[t!]
\centering
\includegraphics[trim=1cm 10cm 3cm 1cm,clip,width=3.5in]{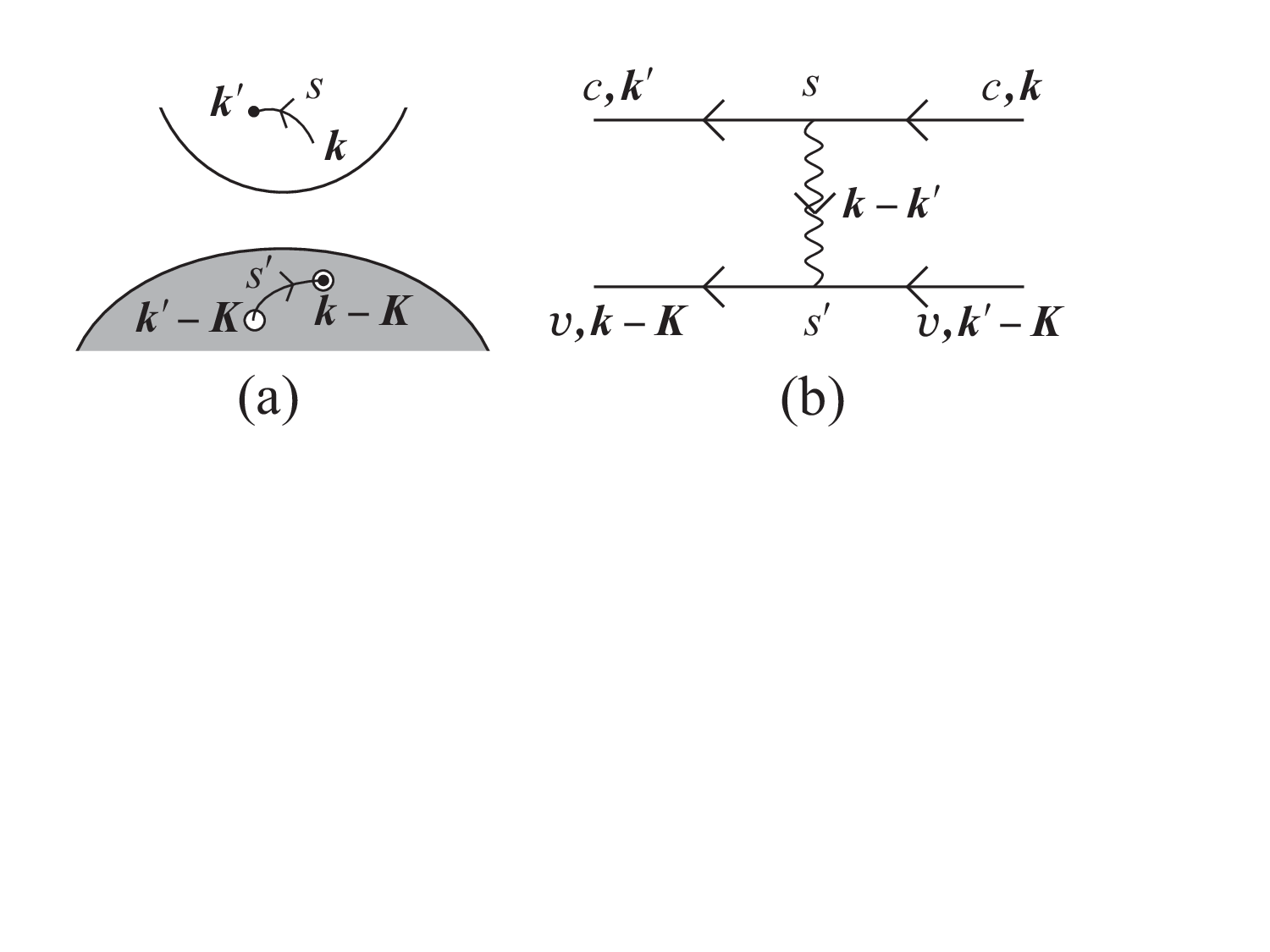}
\vspace{-0.7cm}
\caption{(a) Intraband Coulomb process responsible for the formation of Wannier excitons: Each electron stays in its band. (b) Feynman diagram for this intraband process. Such a process exists whatever the electron spins, $s$ and $s'$.}
\label{fig_3}
\end{figure}

Note that it is possible to achieve the same state change via the \textit{intraband} Coulomb process shown in Fig.~\ref{fig_3}(a) and represented by the Feynman diagram of Fig.~\ref{fig_3}(b). Repetition of these intraband Coulomb processes, that transforms free electron-hole pairs into correlated pairs, are responsible for binding an electron and a hole into a Wannier exciton\cite{Monicbook}.

\begin{figure}[t!]
\centering
\includegraphics[trim=0.2cm 14.3cm 13cm 1cm,clip,width=3.1in]{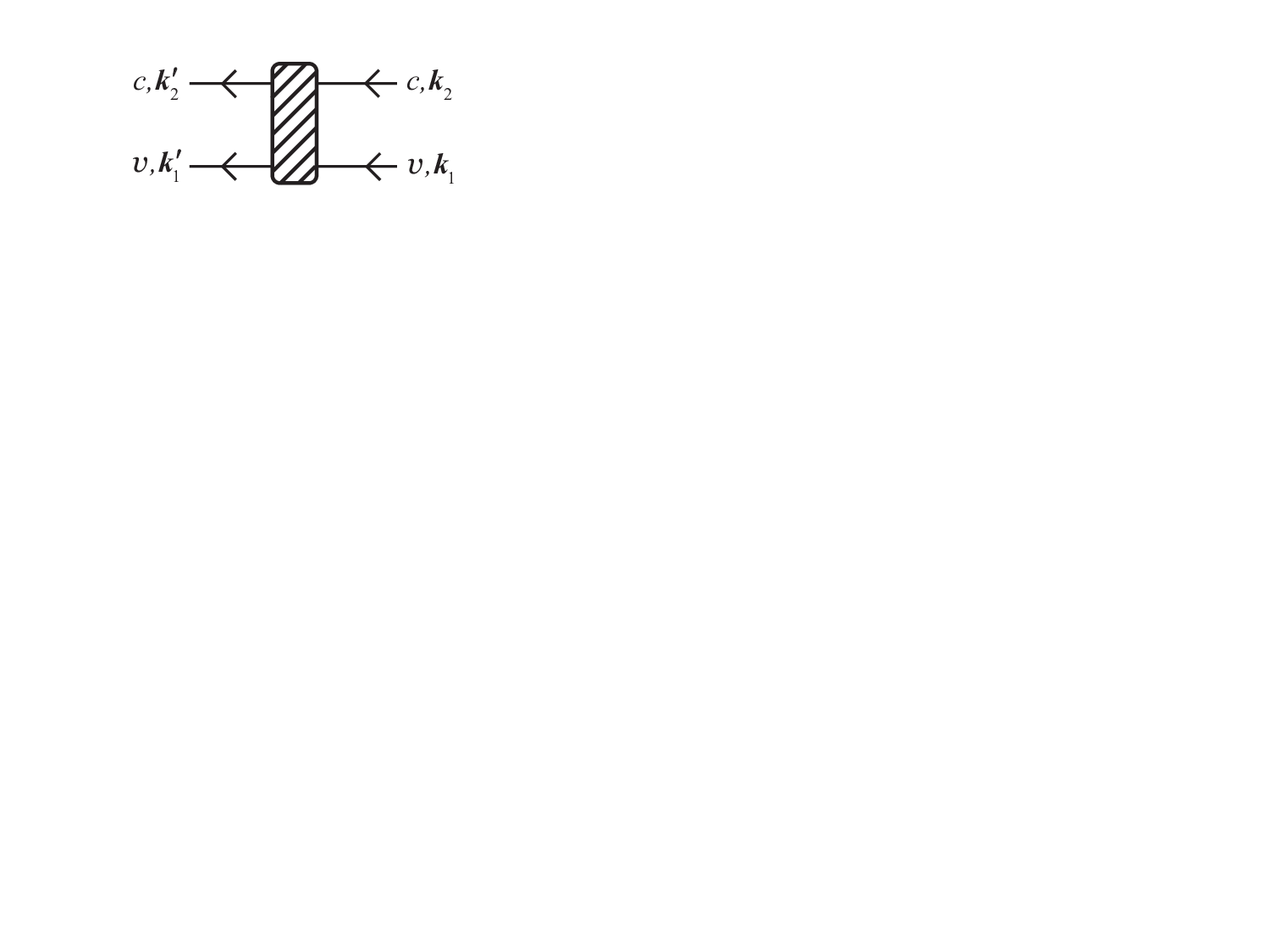}
\vspace{-0.7cm}
\caption{General diagram for scattering between a valence electron $\vk_1$ and a conduction electron $\vk_2$, that keeps the number of electrons in each band. The scatterings that do not keep this number lead to processes far away in energy.}
\label{fig_4}
\end{figure}

\subsection{Generalities on Feynman diagrams}

Feynman diagrams provide an enlightening way to visualize interactions between particles. We here draw them in an unconventional way, with arrows from right to left, in order to match the processes they represent: Indeed, the latter are written with the initial-state destruction operators at the right of final-state creation operators. 

\subsubsection{In terms of valence and conduction electrons}

We consider the Coulomb scattering, shown in Fig.~\ref{fig_4}, in which a pair of valence and conduction electrons changes from the $\left[(v,\vk_1);(c,\vk_2)\right]$ states to the $\left[(v,\vk'_1);(c,\vk'_2)\right]$ states, with $\vk_1+\vk_2=\vk'_1+\vk'_2$ as required by wave vector conservation. Two fundamentally different processes can take place.

\begin{figure}[t!]
\centering
\includegraphics[trim=0.5cm 10.3cm 3.5cm 1cm,clip,width=3.5in]{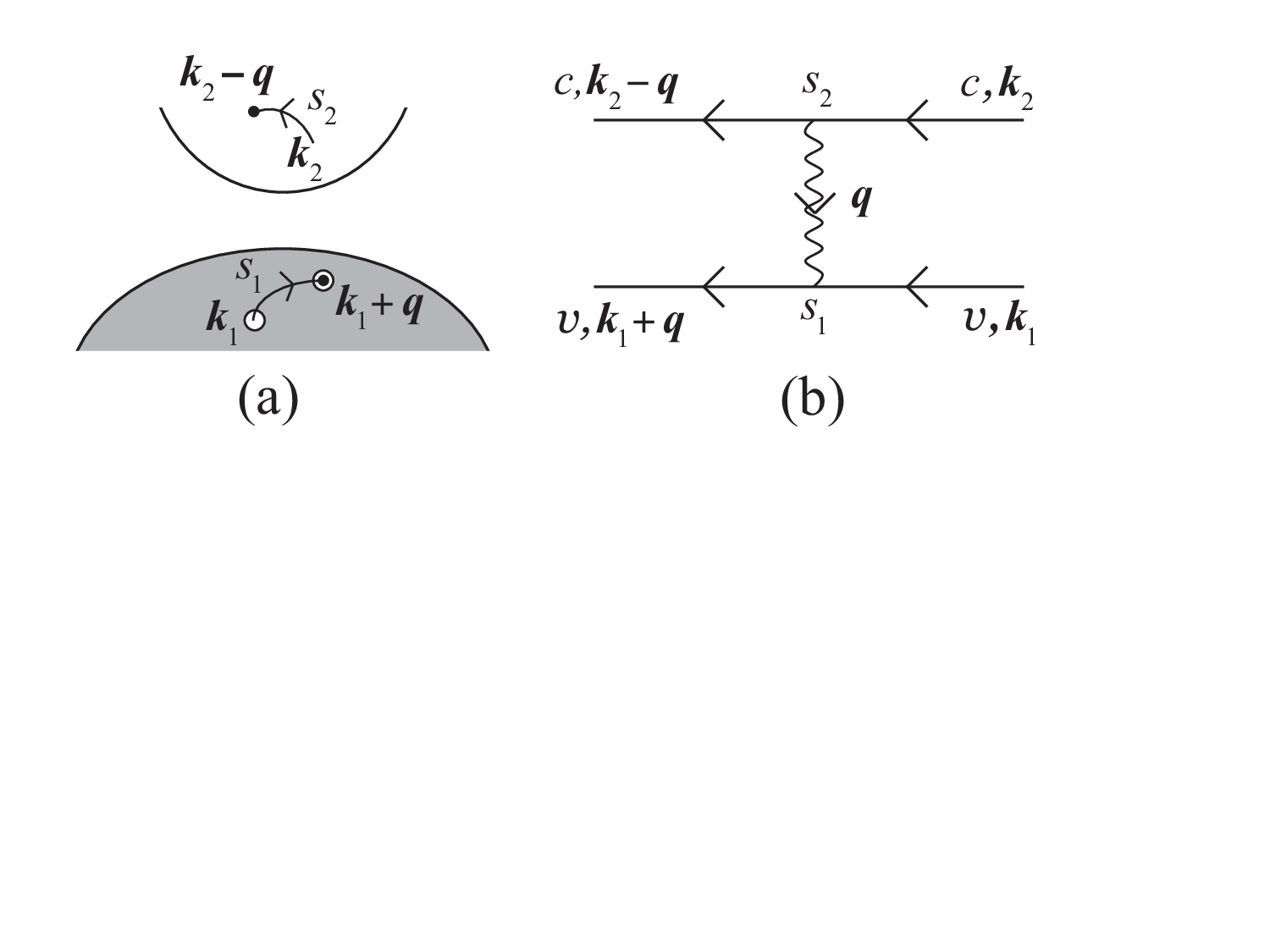}
\vspace{-0.7cm}
\caption{ Figure \ref{fig_4} in the case of \textit{intraband} Coulomb process, that is, conduction and valence electrons staying in their band.}
\label{fig_5}
\end{figure}
\begin{figure}[t!]
\centering
\includegraphics[trim=0.5cm 10.3cm 3.5cm 1cm,clip,width=3.5in]{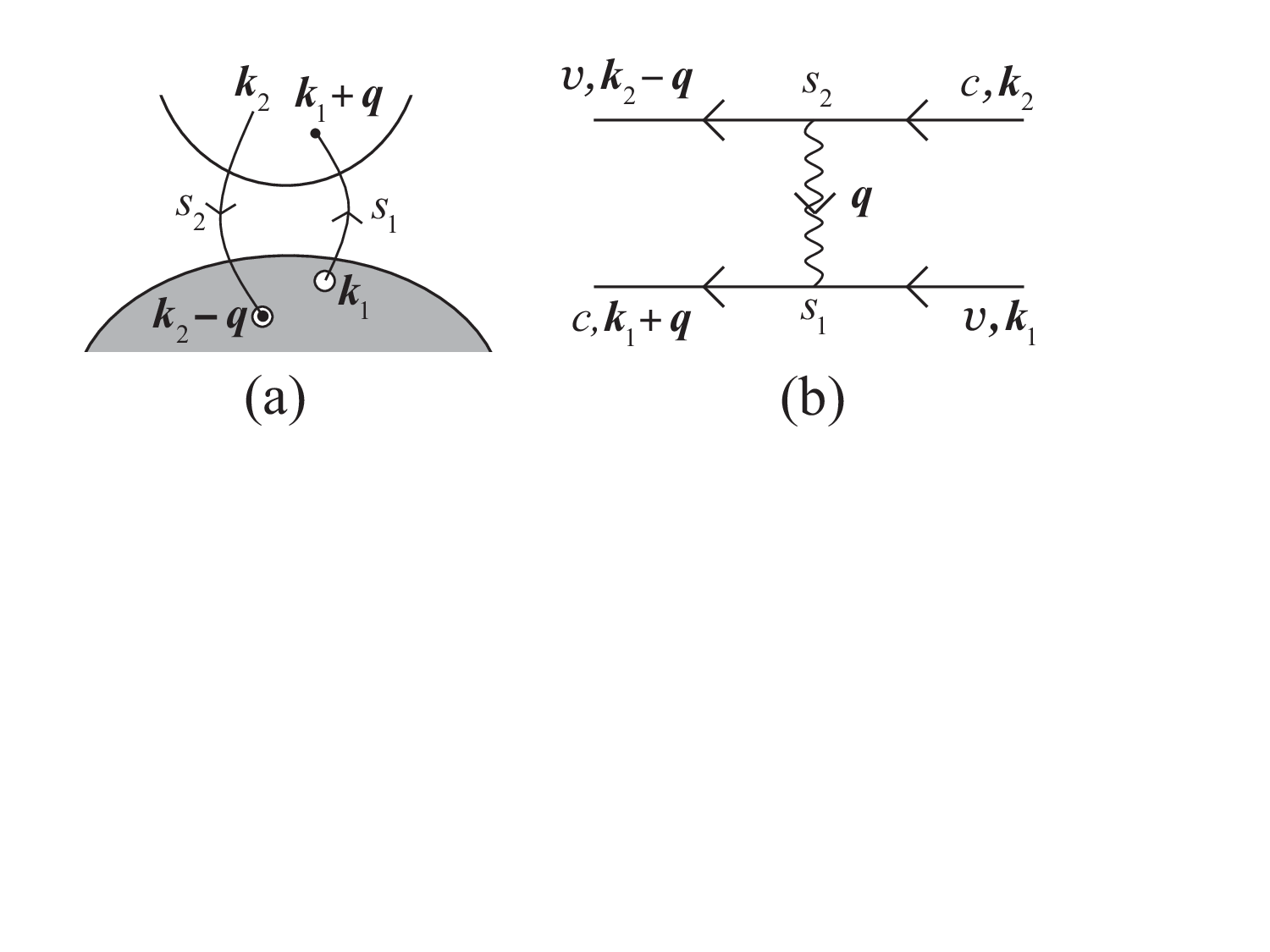}
\vspace{-0.7cm}
\caption{Figure \ref{fig_4} in the case of \textit{interband} Coulomb process: The conduction electron $\vk_2$ fills the empty valence state $\vk_2-\vq$, while the valence electron $\vk_1$  jumps into the conduction state $\vk_1+\vq$. }
\label{fig_6}
\end{figure}

\noindent (1) Each electron stays in its band. This \textit{intraband} Coulomb process for ($\vk'_1, \vk'_2$) written as ($\vk_1+\vq,\vk_2-\vq$) (see Fig.~\ref{fig_5}(a)), is represented by the Feynman diagram of Fig.~\ref{fig_5}(b). Each electron keeps its spin since the Coulomb interaction does not act on spin.

\noindent (2) Each electron changes band. This \textit{interband} Coulomb process that corresponds to $\vk'_1=\vk_2-\vq$ and $\vk'_2=\vk_1+\vq$ (see Fig.~\ref{fig_6}(a)), is represented by the Feynman diagram of Fig.~\ref{fig_6}(b).

\subsubsection{In terms of electrons and holes}

\begin{figure}[t!]
\centering
\includegraphics[trim=0cm 11.7cm 5.5cm 0cm,clip,width=3.5in]{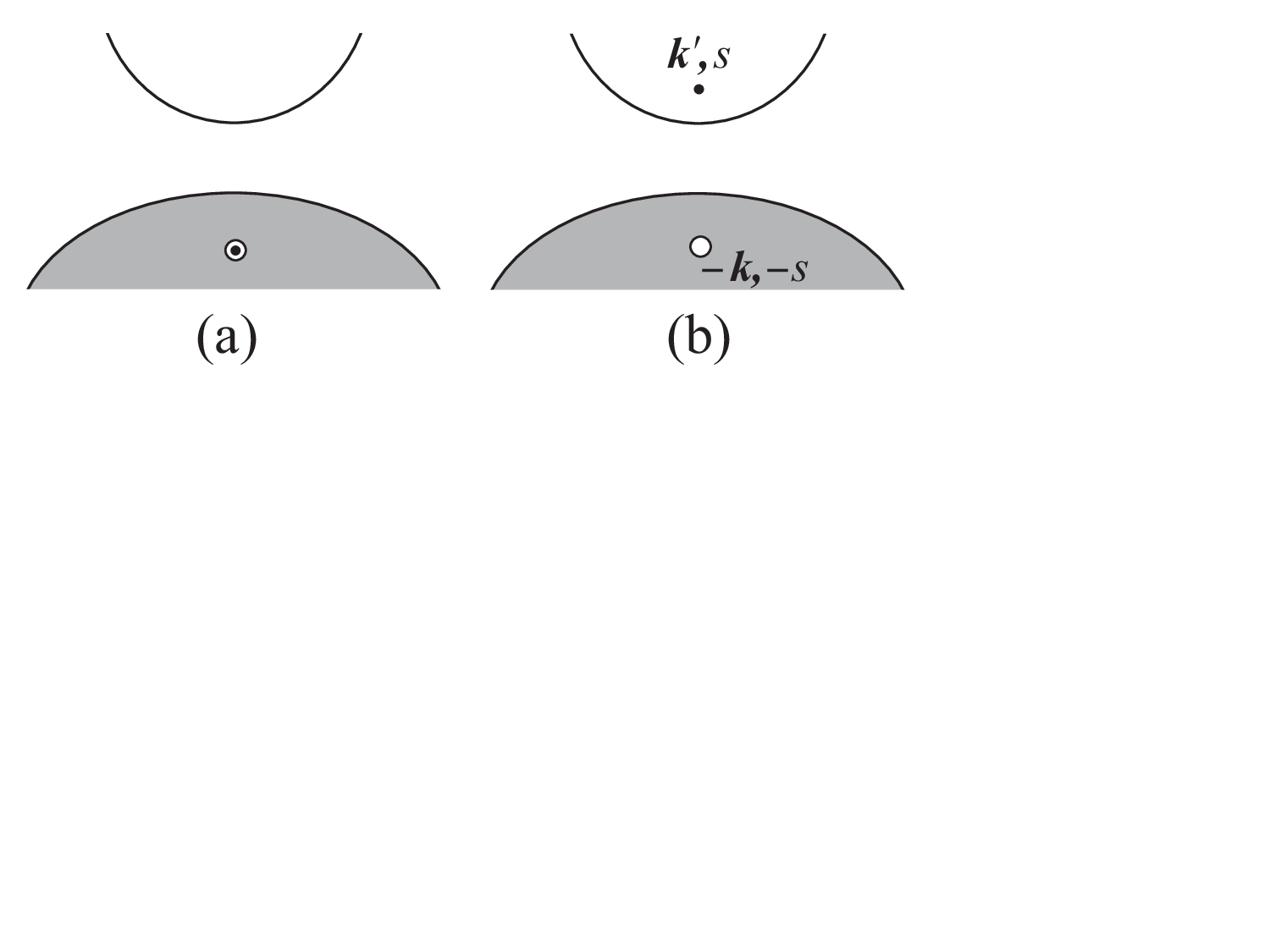}
\vspace{-0.7cm}
\caption{(a) Semiconductor ground state: the conduction band is empty and the valence band is fully occupied; this can be seen as all valence electrons ($\vk,s)$ filling their corresponding valence holes ($-\vk,-s)$. (b) Lowest set of excitations: one valence electron-valence hole pair has ``boiled'': the valence electron $(\vk,s$) now is in the conduction state $(\vk',s)$, while the valence hole $(-\vk,-s$) still is in the valence band. }
\label{fig_7}
\end{figure}

The proper way to handle problems regarding excited semiconductors is not in terms of valence electrons but in terms of valence-electron absences, that is, in terms of electron-hole excitations. This is especially true for interband Coulomb processes. The $(v,\vk_1,s_1)$ electron that jumps to the conduction band, leaves a $(v,\vk_1,s_1)$ empty state in the valence band; this empty state corresponds to a hole with wave vector $-\vk_1$ and spin $-s_1$, on top of the total wave vector and spin of the fully occupied valence band. So, the excitation of a valence electron into the conduction band can be seen as a pair of valence electron-valence hole (see Fig.~\ref{fig_7}(a)) that ``boils'' into a pair of conduction electron-valence hole  (see Fig.~\ref{fig_7}(b)). 

This leads us to replace Fig.~\ref{fig_6} for valence and conduction electrons by Fig.~\ref{fig_8} for electrons and holes. Accordingly, the Feynman diagram of Fig.~\ref{fig_6}(b), that can also be redrawn as in Fig.~\ref{fig_8}(b), appears as in Fig.~\ref{fig_8}(c), when written in terms of electrons and holes, the electron-hole pair wave vector $\vq$ being equal to the wave-vector transfer of the interband Coulomb process. 

The Feynman diagram of Fig.~\ref{fig_1}(a) that represents the interband Coulomb scattering $V_\vK(\vk',\vk)$, is identical to the Feynman diagram of Fig.~\ref{fig_8}(c). A careful handling of the sign change that mathematically appears when turning from valence-electron destruction operator to hole creation operator, as done in Sec.~\ref{sec3G}, further shows that the involved electron-hole pairs with total spin $S_z=0$ are in the spin-singlet state $(S=0,S_z=0)$, not in the singlet-triplet state $(S=1,S_z=0)$ having the same component along $\bf z$.

\section{Coulomb scattering in the Bloch-state basis\label{sec2}}

\subsection{Coulomb interaction in first quantization}

\noindent $\bullet$ The electron-electron Coulomb interaction for $N$ electrons located at $\vr_j$, reads in the first quantization as
\be\label{cheh_ex:2}
V_{e-e}=\frac{1}{2}\sum_{j=1}^{N}\sum_{j'\neq j}^{N} \frac{e^2}{|\vr_j-\vr_{j'}|}
\ee
The good way to tackle this interaction is to turn to the second quantization. An important advantage of this formalism is to avoid writing $N$-electron states as Slater determinants that impose to carefully follow the resulting minus signs induced by fermion exchanges when calculating matrix elements in the $N$-electron subspace. 

 \begin{figure}[t!]
\centering
\includegraphics[trim=1cm 7cm 3cm 0.5cm,clip,width=3.5in]{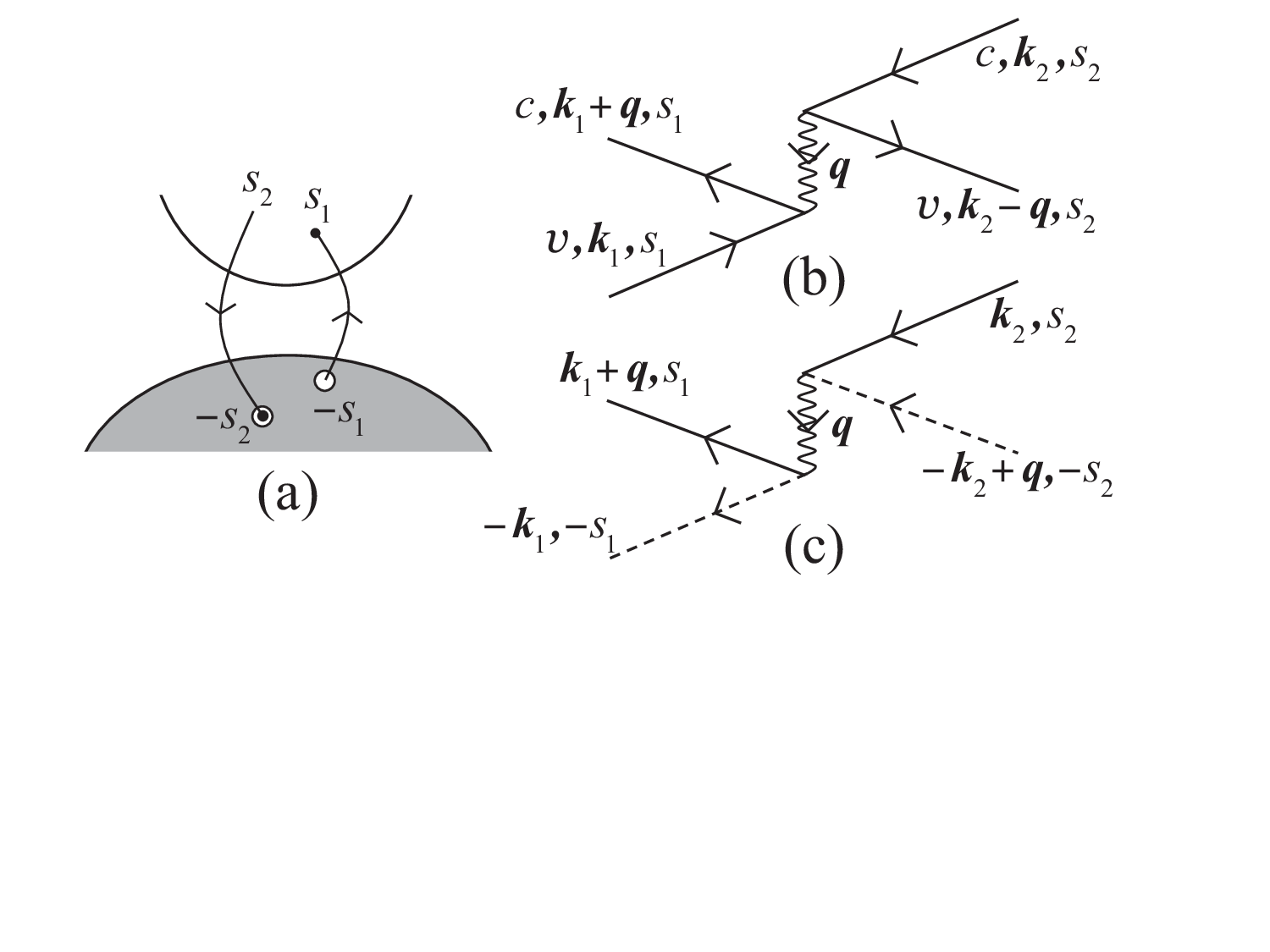}
\vspace{-0.7cm}
\caption{Interband Coulomb process: (a) One electron-hole pair with spins $(s_2,-s_2)$ recombines, while another pair is created. (b) Feynman diagram for the interband process of Fig.~\ref{fig_6}(b), but drawn differently. (c) Same as (b) in terms of electron and hole. }
\label{fig_8}
\end{figure}

The first step in the second quantization is to choose an appropriate one-electron basis. The relevant basis for electrons in a periodic ion lattice hosting Wannier excitons is not made of free electron states $|\vk\ran$ with wave function $\lan \vr|\vk\ran =e^{i\vk\cdot\vr}/L^{3/2}$, for $\vk$ quantized in $2\pi/L$ in a sample volume $L^3$ (within the Born-von Karman boundary conditions), but it is made of Bloch states $|n,\vk\ran$ that differentiate electrons with same wave vector $\vk$ in different bands $n$. According to the Bloch theorem\cite{Kittelbook,Bloch1929,Merminbook}, their wave function
\be\label{cheh_ex:3}
\lan \vr|n,\vk\ran=\frac{e^{i\vk\cdot \vr}}{L^{3/2}}u_{n,\vk}(\vr)\equiv e^{i\vk\cdot \vr}\lan \vr|u_{n,\vk}\ran
\ee
appears as the free electron wave function $\lan \vr|\vk\ran$ modulated by a Bloch function $u_{n,\vk}(\vr)$ that has the lattice periodicity
\be
\label{cheh_ex:4}
 u_{n,\vk}(\vr)=u_{n,\vk}(\vr+\vR_\ell)    
\ee
for any lattice vector $\vR_\ell$ with $\ell=(1,\cdots,N)$. The appropriate way to handle the lattice periodicity is not in terms of these $N$ lattice vectors $\vR_\ell$ but in terms of the $N$ reciprocal vectors $\textbf{G}_m$, by Fourier expanding $u_{n,\vk}(\vr)$ as
\be\label{cheh_ex:5}
u_{n,\vk}(\vr)=\sum_{m=0}^{N-1} e^{i\textbf{G}_m\cdot\vr} \,u_{n,\vk;\textbf{G}_m}
\ee
for $\textbf{G}_m$ quantized in $2\pi/a_c$ with $a_c$ being the lattice cell size,
 in order to have $e^{i\textbf{G}_m\cdot\vR_\ell}=1$ whatever $\vR_\ell$.

\noindent $\bullet$ The Bloch states are eigenstates of a one-body Hamiltonian $h$, 
\be\label{cheh_ex:5_1}
0=(h-\va_{n,\vk})|n,\vk\ran
\ee
 which corresponds to a free electron with mass $m_0$ that interacts with all the lattice ions, and also with all the other electrons through a one-body average repulsive interaction that has the lattice periodicity\cite{Monicbook}. The $h$ Hamiltonian in addition contains a constant term  that comes from the ion-ion interaction, as required for $h$ to represent a neutral system, in order to avoid spurious overextensive terms\cite{Monicbook}.  
  
 This $h$ Hamiltonian follows from the system Hamiltonian $H$ for $N$ electrons labeled by $j$ and $N$ ions with charge $|e|$ and infinite mass, located at $\vR_\ell$, 
\bea\label{cheh_ex:5_2}
H&=&\sum_{j=1}^N\frac{\vp^2_j}{2m_0}-\sum_{j=1}^N\sum_{\ell=1}^N\frac{e^2}{|\vr_j-\vR_\ell|}
 \\
&&
+\frac{1}{2}\sum_{j=1}^N\sum_{j'\not=j}\frac{e^2}{|\vr_j-\vr_{j'}|}
+\frac{1}{2}\sum_{\ell=1}^N\sum_{\ell'\not=\ell}\frac{e^2}{|\vR_\ell-\vR_{\ell'}|}
\nn 
\eea
that we rewrite as
\be\label{cheh_ex:5_3}
H=\sum_{j=1}^Nh_j +\mathcal{V}_{Coul}
\ee

This splitting concentrates the two-body part of the problem, \textit{i.e.}, the one leading to many-body effects, into 
\be
\label{9}
\mathcal{V}_{Coul}=V_{e-e}-\overline{V}_{e-e}
\ee
with $V_{e-e}$  given in Eq.~(\ref{cheh_ex:2}), while 
\bea
\label{9'}
\overline{V}_{e-e}=\sum_{j=1}^N \overline{v}_{e-e}(\vr_j)
\eea
with $\overline{v}_{e-e}(\vr_j)=\overline{v}_{e-e}(\vr_j+\vR_\ell)$ whatever $\vR_\ell$, is the one-body average electron-electron Coulomb interaction that is introduced to properly define the Bloch-state basis.

Another important requirement for this average interaction is to catch most of the electron-electron repulsion in order to possibly treat $\mathcal{V}_{Coul}$ in a perturbative way when dealing with many-body effects.  The simplest choice is to replace the electron gas by a jellium\cite{Haug&Koch,Mahan} having the same charge density, $Ne/L^3$. Its interaction with the $\vr_j$ electron would then lead to
 \be
\label{9'_1}
\overline{v}^{(jel)}_{e-e}(\vr_j)=\frac{1}{2}
\int_{L^3}\frac{d^3r}{L^3} 
\frac{N e^2}{|\vr_j-\vr|}
\ee
the $1/2$ prefactor coming from the fact that the $\vr_j$ electron belongs to the jellium. Note that this $\overline{v}^{(jel)}_{e-e}(\vr_j)$ interaction has the required lattice periodicity in the large $L$ limit.

The resulting one-body Hamiltonian, from which the Bloch states are constructed, then reads 
\bea
\label{10}
h_j=\frac{\vp^2_j}{2m_0} +v(\vr_j) 
 \eea
 with the total Coulomb interaction given by
  \bea
  \label{10'}
v(\vr_j)
=-\sum_{\ell=1}^N\frac{e^2}{|\vr_j-\vR_\ell|}
+\overline{v}_{e-e}(\vr_j)\hspace{2cm}
\\
+\frac{1}{2N}\sum_{\ell=1}^N\sum_{\ell'\not=\ell}\frac{e^2}{|\vR_\ell-\vR_{\ell'}|}
\nn 
 \eea
so that  $v(\vr_j)=v(\vr_j+\vR_\ell)$ whatever $\vR_\ell$, as required.

\subsection{Electron-electron interaction using Bloch states}

\noindent $\bullet$ Following the second quantization procedure\cite{Monicbook}, we can write the two-body electron-electron interaction  $V_{e-e}$ given in Eq.~(\ref{cheh_ex:2}) in terms of the creation operators $\hat{a}^\dag_{n,\vk,s}$ for electrons with spin $s$ in the Bloch state $(n,\vk)$. The $V_{e-e}$ interaction then leads to the operator
\bea \label{cheh_ex:6}
\widehat{V}_{e-e}=\frac{1}{2}\sum_{\vq_1\vq_2}\sum_{\{n,\vk,s\}}\!\!\! V\left(\begin{smallmatrix}
n'_2,\vk_2-\vq_2& \, n_2,\vk_2\\ n'_1,\vk_1+\vq_1& \, n_1,\vk_1\end{smallmatrix}\right)\hspace{2cm}\\ 
\hat{a}^\dag_{n'_1,\vk_1+\vq_1,s_1}\hat{a}^\dag_{n'_2,\vk_2-\vq_2,s_2}\hat{a}_{n_2,\vk_2,s_2} \hat{a}_{n_1,\vk_1,s_1}\nn
\eea
since the Coulomb interaction does not act on spin. The scattering amplitude, here written as a square box to evidence that the spin-$s_1$ electron in the Bloch state $(n_1,\vk_1)$ ends with the same spin in the Bloch state $(n'_1,\vk_1+\vq_1)$, reads in terms of the Bloch-state wave functions (\ref{cheh_ex:3}) as
\bea  \label{cheh_ex:7}
\lefteqn{V\left(\begin{smallmatrix}
n'_2,\vk_2-\vq_2
\,\, \,  n_2,\vk_2\\ n'_1,\vk_1+\vq_1
 \,\,\,   n_1,\vk_1\end{smallmatrix}\right)=\iint_{L^3} d^3r_1 d^3r_2}
 \,\,\,\,\,\,\,\,\,\,\,\,\,\,\,\,\,\,\,\,\,\,\,\,\,\,\,\,\,\,\,\,
  \,\,\,\,\,\,\,\,\,\,\,\,\,\,\,\,\,\,\,\,\,\,\,\,\,\,\,\,\,\,\,\, \,\,\,\,\,\,\,\,\,\,\,\,\,\,\,\,\,\,\,\,\,\,\,\,\,\,\,\,\,\,\,\, \,\,\,\,\,\,\,\,\,\,\,\,\,\,\,\,\,\,\,\,\,\,\,\,\,\,\,\,\,\,\,\,\,\,\,\,\,\,\,\,\,\,\,\,\,\,\,\,\,\,\,\,
 \\
 \lan n'_1,\vk_1{+}\vq_1|\vr_1\ran  \lan n'_2,\vk_2{-}\vq_2|\vr_2\ran \frac{e^2}{|\vr_1{-}\vr_2|}\lan \vr_2|n_2,\vk_2\ran  \lan \vr_1|n_1,\vk_1\ran\nn
\eea

\noindent $\bullet$ To calculate this quantity, we write $\vr$ as $\vR_\ell+ \boldsymbol{\rho}$; this divides the $\vr$ integrals over the sample volume $L^3$ into a sum over the $N$ lattice vectors $\vR_\ell$ and an integral over a unit cell volume $a_c^3$, as
\be\label{cheh_ex:8}
\int_{L^3}d^3r=\sum_{\vR_\ell}\int_{a_c^3}d^3\rho
\ee
Because of the  lattice periodicity (\ref{cheh_ex:4}) of the Bloch functions, we can rewrite Eq.~(\ref{cheh_ex:7}),  for $\vr_2=\boldsymbol{\rho}_2+\vR_{\ell_2}$ and $\vr_1=\boldsymbol{\rho}_1+\vR_{\ell_1}=\boldsymbol{\rho}_1+\vR_{\ell}+\vR_{\ell_2}$, as 
\bea
\label{14}  
V\left(\begin{smallmatrix}
n'_2,\vk_2-\vq_2&\, n_2,\vk_2\\ n'_1,\vk_1+\vq_1&\, n_1,\vk_1\end{smallmatrix}\right)=\frac{1}{L^6} \iint_{a_c^3} d^3\rho_1 d^3\rho_2  \, e^{i(\vq_2\cdot\boldsymbol{\rho}_2-\vq_1\cdot \boldsymbol{\rho}_1)}\nn\\
 u^\ast_{n'_1,\vk_1+\vq_1}(\boldsymbol{\rho}_1) \,  u^\ast_{n'_2,\vk_2-\vq_2}(\boldsymbol{\rho}_2)\,u_{n_2,\vk_2}(\boldsymbol{\rho}_2)\, u_{n_1,\vk_1}(\boldsymbol{\rho}_1)\nn\\
 \sum_{\ell=1}^N e^{-i\vq_1\cdot\vR_\ell}\frac{e^2}{|\vR_\ell+\boldsymbol{\rho}_1-\boldsymbol{\rho}_2 |}\sum_{{\ell_2}=1}^N  e^{i(\vq_2-\vq_1)\cdot\vR_{\ell_2}}\hspace{0.3cm}\label{cheh_ex:9}
\eea
Since the sum over ${\ell_2}$ is equal to $N\delta_{\vq_1,\vq_2}$ for \textbf{q} quantized in $2\pi /L$ with $L^3= N a_c^3$, we recover the expected wave-vector conservation for the scattered electron pair.
 
This gives the electron-electron interaction (\ref{cheh_ex:6}) in the Bloch-state basis as
 \bea \label{17a}
\widehat{V}_{e-e}=\frac{1}{2} \sum_\vq    \sum_{\{n,\vk,s\}}\!\!\! V\left(\begin{smallmatrix}
n'_2,\vk_2-\vq& \, n_2,\vk_2\\ n'_1,\vk_1+\vq& \, n_1,\vk_1\end{smallmatrix}\right)\hspace{2cm}\\ 
\hat{a}^\dag_{n'_1,\vk_1+\vq,s_1}\hat{a}^\dag_{n'_2,\vk_2-\vq,s_2}\hat{a}_{n_2,\vk_2,s_2} \hat{a}_{n_1,\vk_1,s_1}\nn
\eea
 with the Coulomb scattering amplitude given by
\bea
\label{17b}  
V\left(\begin{smallmatrix}
n'_2,\vk_2-\vq&\, n_2,\vk_2\\ n'_1,\vk_1+\vq&\, n_1,\vk_1\end{smallmatrix}\right)=\frac{1}{N} \iint_{a_c^3}  \frac{d^3\rho_1}{a_c^3}          \frac{d^3\rho_2}{a_c^3}  \, e^{i\vq\cdot (\boldsymbol{\rho}_2- \boldsymbol{\rho}_1)}\nn\\
 u^\ast_{n'_1,\vk_1+\vq}(\boldsymbol{\rho}_1) \,  u^\ast_{n'_2,\vk_2-\vq}(\boldsymbol{\rho}_2)\,u_{n_2,\vk_2}(\boldsymbol{\rho}_2)\, u_{n_1,\vk_1}(\boldsymbol{\rho}_1)\nn\\
 \sum_{\ell=1}^N e^{-i\vq\cdot\vR_\ell}\frac{e^2}{|\vR_\ell+\boldsymbol{\rho}_1-\boldsymbol{\rho}_2 |}
\eea

 \subsection{Average electron-electron interaction}

 \noindent $\bullet$ We can also write the one-body average electron-electron interaction $\overline{V}_{e-e}$ as an operator in the Bloch state basis. Using the second quantization procedure, this one-body operator first appears as
 \be
 \label{18}  \widehat{\overline{V}}_{e-e}=\sum_s\sum_{n',n}\sum_{\vk,\vq}\overline{v}_{e-e}(n',\vk+\vq;n,\vk)\,\hat{a}^\dag_{n',\vk+\vq,s}\hat{a}_{n,\vk,s}
 \ee
 with the prefactor given by
 \be \label{18_1}
 \overline{v}_{e-e}(n',\vk+\vq;n,\vk)=\int_{L^3}d^3r \lan n',\vk+\vq|\vr\ran \overline{v}_{e-e}(\vr)\lan \vr|n,\vk\ran
 \ee
 To calculate this prefactor, we use the Bloch-state wave functions (\ref{cheh_ex:3}), and we split the integral over $\vr$  according to Eq.~(\ref{cheh_ex:8}). The above scattering amplitude then reads, for $\vr=\vR_\ell+\boldsymbol{\rho}$, as
 \bea \label{18_2}
 \overline{v}_{e-e}(n',\vk+\vq;n,\vk)=\frac{1}{L^3} \sum_{\ell=1}^N\int_{a_c^3}d^3\rho\, e^{-i\vq\cdot(\vR_\ell+\boldsymbol{\rho})}\hspace{1cm}\\
 u^\ast_{n',\vk+\vq}(\boldsymbol{\rho})\,u_{n,\vk}(\boldsymbol{\rho})\,\overline{v}_{e-e}(\vR_\ell+\boldsymbol{\rho})\nn
 \eea
Since $\overline{v}_{e-e}(\vR_\ell+\boldsymbol{\rho})=\overline{v}_{e-e}(\boldsymbol{\rho})$ is periodic, the sum over $\ell$ reduces to $\sum_{\ell}e^{-i\vq\cdot\vR_\ell}=N\delta_{\vq,\bf0}$; so, we get
\begin{subeqnarray} \label{18_3}
&&\overline{v}_{e-e}(n',\vk+\vq;n,\vk)=\delta_{\vq,\bf0}\,\overline{v}_\vk(n',n)\hspace{1.5cm}\\
&&\overline{v}_\vk(n',n)= \int_{a_c^3}\frac{d^3\rho}{a_c^3} \,u^\ast_{n',\vk}(\boldsymbol{\rho})\,\overline{v}_{e-e}(\boldsymbol{\rho})\, u_{n,\vk}(\boldsymbol{\rho}) 
\end{subeqnarray}

By using these results, we find that the one-body average electron-electron interaction (\ref{18}) reduces to
 \be
 \label{18_4}  \widehat{\overline{V}}_{e-e}=\sum_{n',n}\sum_{\vk, s}\overline{v}_\vk(n',n)\,\hat{a}^\dag_{n',\vk,s}\hat{a}_{n,\vk,s}
 \ee
 It allows a band change $(n'\not=n)$, but not a wave-vector transfer.

\noindent $\bullet$ It is possible to rewrite this one-body operator as a two-body operator by noting that $\hat{\textrm{I}}^{(N)}$ defined as
 \be
  \label{19}
\hat{\textrm{I}}^{(N)}=\frac{1}{N}\sum_{n,\vk, s}\hat{a}^\dag_{n,\vk,s}\hat{a}_{n,\vk,s}=\frac{1}{N}\sum_{n',n}\sum_{\vk, s}\delta_{n',n}\hat{a}^\dag_{n',\vk,s}\hat{a}_{n,\vk,s}
 \ee
 with $\delta_{n',n}$ possibly written as
 \be
 \delta_{n',n}=\lan n',\vk|n,\vk\ran=\int_{a_c^3}\frac{d^3\rho}{a_c^3}u^\ast_{n',\vk}(\boldsymbol{\rho})\,u_{n,\vk}(\boldsymbol{\rho})
 \ee
reduces to the identity operator when acting on any $N$-electron state in the Bloch-state basis. 
 
Consequently, the operator $\hat{a}^\dag_{n',\vk,s} \hat{a}_{n,\vk,s}$ in Eq.~(\ref{18_4}) acts on such $N$-electron state in the same way as $\hat{a}^\dag_{n',\vk,s} \hat{\textrm{I}}^{(N-1)} \hat{a}_{n,\vk,s}$. So, we can rewrite $ \widehat{\overline{V}}_{e-e}$ in Eq.~(\ref{18}) as
 \bea
  \label{20} 
  \widehat{\overline{V}}_{e-e}=\frac{1}{2}\sum_{\{n',n\}}\sum_{\{\vk,s\}} \overline{V}\left(\begin{smallmatrix}
n'_2,\vk_2&\, n_2,\vk_2\\ n'_1,\vk_1&\, n_1,\vk_1\end{smallmatrix}\right)\hspace{1cm}\\
\hat{a}^\dag_{n'_1,\vk_1,s_1}\hat{a}^\dag_{n'_2,\vk_2,s_2}\hat{a}_{n_2,\vk_2,s_2}\hat{a}_{n_1,\vk_1,s_1}\nn
 \eea
 the scattering amplitude for $N-1\sim N$ large, being given by
 \bea\label{20_1} 
 \overline{V}\left(\begin{smallmatrix}
n'_2,\vk_2&\, n_2,\vk_2\\ n'_1,\vk_1&\, n_1,\vk_1\end{smallmatrix}\right)=\hspace{4.5cm}\\
\frac{1}{N}\iint_{a_c^3}\frac{d^3\rho_1}{a_c^3}\frac{d^3\rho_2}{a_c^3}\,
\Big(\overline{v}_{e-e}(\boldsymbol{\rho}_1)+\overline{v}_{e-e}(\boldsymbol{\rho}_2)\Big)\nn\\
u^\ast_{n'_1,\vk_1}(\boldsymbol{\rho}_1)\,u^\ast_{n'_2,\vk_2}(\boldsymbol{\rho}_2)\,u_{n_2,\vk_2}(\boldsymbol{\rho}_2)\,u_{n_1,\vk_1}(\boldsymbol{\rho}_1)\nn
 \eea
 The main characteristic of this average electron-electron interaction is to be associated with zero-wave-vector transfers.

  \subsection{Semiconductor Coulomb interaction $\mathcal{V}_{Coul}$}
 
 \noindent $\bullet$ When used into the semiconductor Coulomb interaction $\mathcal{V}_{Coul}$ defined in Eq.~(\ref{9}), we find that this interaction is represented in the Bloch-state basis by a two-body operator, with a scattering that splits in a natural way into a part associated with zero-wave-vector transfers that comes from $V_{e-e}-\overline{V}_{e-e}$ and a part associated with finite-wave-vector transfers that only comes from $V_{e-e}$ 
  \bea \label{23a}
\widehat{\mathcal{V}}_{Coul}=\frac{1}{2} \sum_\vq    \sum_{\{n,\vk,s\}}\!\!\!
 \mathcal{V}\left(\begin{smallmatrix}
n'_2,\vk_2-\vq& \, n_2,\vk_2\\ n'_1,\vk_1+\vq& \, n_1,\vk_1\end{smallmatrix}\right)\hspace{2cm}\\ 
\hat{a}^\dag_{n'_1,\vk_1+\vq,s_1}\hat{a}^\dag_{n'_2,\vk_2-\vq,s_2}\hat{a}_{n_2,\vk_2,s_2} \hat{a}_{n_1,\vk_1,s_1}\nn
\eea
 with the scattering amplitude given by
 \bea
  \label{22}
 \mathcal{V}\left(\begin{smallmatrix}
n'_2,\vk_2-\vq& \, n_2,\vk_2\\ n'_1,\vk_1+\vq& \, n_1,\vk_1\end{smallmatrix}\right)
= V\left(\begin{smallmatrix}
n'_2,\vk_2-\vq& \, n_2,\vk_2\\ n'_1,\vk_1+\vq& \, n_1,\vk_1\end{smallmatrix}\right)\,\,\,\,\,\,\,\,\,\,\,\,\,\,\,\,\,\,\,\,\,\,\,\,\,\,\,\,\,\,\,\,\,\,\,\,
\\ \nn
- \delta_{\vq,\v0}
\overline{V}\left(\begin{smallmatrix}
n'_2,\vk_2& \, n_2,\vk_2\\ n'_1,\vk_1& \, n_1,\vk_1\end{smallmatrix}\right)\hspace{1cm}
 \eea

 \noindent $\bullet$ By properly choosing the average electron-electron interaction, it is possible to exactly cancel the singular $\vq=\bf0$ part of the  $\widehat{V}_{e-e}$ scattering, as required to ultimately treat $\mathcal{V}_{Coul}$ in a perturbative way. The operator $\widehat{\mathcal{V}}_{Coul}$ then reduces to the part of $\widehat{V}_{e-e}$ for finite wave-vector transfers only.

\subsection{Interband scattering}

\noindent $\bullet$ The $\widehat{V}_{e-e}$ Coulomb interaction (\ref{cheh_ex:6}) generates two types of  terms: The ones $(n'_1=c,n_1=c; n'_2=v,n_2=v)$ and $(n'_1=v, n_1=v; n'_2=c, n_2=c)$ are associated with \textit{intraband} processes, while the ones $(n'_1=c,n_1=v; n'_2=v,n_2=c)$ and $(n'_1=v, n_1=c; n'_2=c, n_2=v)$ are associated with \textit{interband} processes.  These two pairs of terms remove the $1/2$ prefactor in the $\widehat{V}_{e-e}$ Coulomb interaction given in Eq.~(\ref{cheh_ex:6}).

 In the following, we will focus on the interband Coulomb interaction. It appears in terms of valence and conduction electron operators as 
\bea\label{cheh_ex:12}
\widehat{V}^{(inter)}_{e-e}=\sum_\vK\sum_{s',s} \sum_{\vk',\vk} V_\vK(\vk',\vk)\hspace{2.8cm}\\
\hat{a}^\dag_{c,\vk',s'} 
\hat{a}^\dag_{v,\vk-\vK,s}  \hat{a}_{c,\vk,s} \hat{a}_{v,\vk'-\vK,s'}\nn
\eea
The interband Coulomb scattering, shown in  Fig.~\ref{fig_1}(b), is given, according to Eq.~(\ref{cheh_ex:9}), by 
\be
\label{cheh_ex:14}
V_\vK(\vk',\vk)\equiv V\left(\begin{smallmatrix} v,\vk-\vK& \,c,\vk\\ c,\vk' & \, v,\vk'-\vK \end{smallmatrix}\right)=\sum_{\ell=1}^N V_\vK(\vk',\vk; \vR_\ell)
\ee
with, by rewriting $(\boldsymbol{\rho}_1, \boldsymbol{\rho}_2)$ as $(\boldsymbol{\rho}', \boldsymbol{\rho})$, $\vq$ as $\vK$ and $(\vk_2,\vk_1)$ as $(\vk, \vk'-\vK)$ in Eq.~(\ref{17b}),
\bea
\label{17}
V_\vK(\vk',\vk; \vR_\ell)
=\frac{1}{  N  }
\iint_{a_c^3} \frac{d^3\rho'}{a_c^3} \frac{d^3\rho}{a_c^3} 
 \,\,\,\,\, \,\,\,\,\, \,\,\,\,\, \,\,\,\,\,\,\,\,\,\,\,\,\,\,\,\,\,\,\,\,\,\,\,\,\,\,\,\,\,\,\,\,\,\,\,\,\,\,\,\,
\\
 e^{-i\vK\cdot(\vR_\ell+\boldsymbol{\rho'}-\boldsymbol{\rho})}
 \frac{e^2}{|\vR_\ell+\boldsymbol{\rho'}-\boldsymbol{\rho} |}
 \,\,
 w^\ast_{\vK,\vk'}(\boldsymbol{\rho}')w_{\vK,\vk}(\boldsymbol{\rho})
  \nn
\eea
for $w_{\vK,\vk}(\boldsymbol{\rho})$ defined as
\be
\label{5"}
w_{\vK,\vk}(\boldsymbol{\rho})\equiv u^\ast_{v,\vk-\vK}(\boldsymbol{\rho})\,u_{c,\vk}(\boldsymbol{\rho})
\ee

\noindent $\bullet$ There are two ways to calculate the interband scattering amplitude $V_\vK(\vk',\vk)$.

\noindent (\textit{i}) The usual way is to stay in real space and to tackle the $\vR_\ell$ sum in Eq.~(\ref{cheh_ex:14}) by isolating its $\vR_\ell=\textbf{0}$ term, identified with the so-called ``short-range'' Coulomb process, from the $\vR_\ell \neq \textbf{0}$ sum associated with ``long-range'' processes.

\noindent (\textit{ii}) The other way, which is the appropriate one when  dealing with periodic systems, is to handle the lattice periodicity of the Bloch functions by expanding these functions on reciprocal vectors $\textbf{G}_m$ according to Eq.~(\ref{cheh_ex:5}). Equation (\ref{cheh_ex:14}) then appears as
 \bea
\label{17'}
V_\vK(\vk',\vk)=\sum_{m=0}^{N-1} V_{\vK,\textbf{G}_m}(\vk',\vk)
\eea

While the latter way is mathematically appropriate for periodic systems, both ways, that indeed produce the same result, are of interest because they bring out different aspects of the problem. In particular, the singularity, in the $\vK\rightarrow \bf0$ limit, of the single $V_{\vK,\textbf{G}_0=\v0}(\vk',\vk)$ term, is the same as the singularity that appears in the sum of $ V_\vK(\vk',\vk; \vR_\ell)$ over \textit{all} lattice vectors $\vR_\ell \neq \textbf{0}$, associated with long-range Coulomb processes.

\subsection{Formulation in the reciprocal space}

\noindent $\bullet$  To rewrite the interband Coulomb scattering amplitude $V_\vK(\vk',\vk)$ defined in Eq.~(\ref{cheh_ex:14}) as a sum over reciprocal vectors $\textbf{G}_m$ instead of lattice vectors $\vR_\ell$, we first note that since  $u_{n,\vk}(\boldsymbol{\rho})$'s are periodic functions, we can expand their product defined in Eq.~(\ref{5"}) as
\bea
\label{5'}
w_{\vK,\vk}(\boldsymbol{\rho})
=\sum_{m=0}^{N-1} e^{i\textbf{G}_m\cdot \boldsymbol{\rho}} \,w_{\vK,\vk;\textbf{G}_m}
\eea
with the prefactor given by
\be\label{cheh_ex:25}
w_{\vK,\vk;\textbf{G}_m}
=\lan u_{v,\vk-\vK}|e^{-i\textbf{G}_m\cdot\vr}|u_{c,\vk}\ran
\ee
To check it, we use Eqs.~(\ref{cheh_ex:3},\ref{cheh_ex:8}). The above RHS then reads, for $\vr=\boldsymbol{\rho} + \vR_\ell$ and $e^{-i\textbf{G}_m\cdot\vR_\ell}=1$ whatever $\vR_\ell$, as
\bea
\Big( \sum_{\vR_\ell}\int_{a_c^3}d^3\rho  \Big)
\,\,  \frac{u^\ast_{v,\vk-\vK}(\vr)}{L^{3/2}}  \,\, e^{-i\textbf{G}_m\cdot\vr}  \,\, \frac{u_{c,\vk}(\vr)}{L^{3/2}}
\nn
\\ 
=\int_{a_c^3}\frac{d^3\rho}{a_c^3}  e^{-i\textbf{G}_m\cdot\boldsymbol{\rho}}       
w_{\vK,\vk}(\boldsymbol{\rho})
 \eea

\noindent $\bullet$ By inserting Eq.~(\ref{5'}) into Eq.~(\ref{17}), the interband Coulomb scattering (\ref{cheh_ex:14}) then appears as
\bea
\label{cheh_ex:16}
V_\vK(\vk',\vk)=\sum_{m'm}   \, W_\vK(\textbf{G}_{m'},\textbf{G}_m) \,
w^\ast_{\vK,\vk',\textbf{G}_{m'}}  \, w_{\vK,\vk;\textbf{G}_m}
\eea
with the Coulomb part concentrated into
\bea\label{cheh_ex:17}
W_\vK(\textbf{G}_{m'},\textbf{G}_m)=
\iint_{a_c^3}\frac{d^3\rho'}{a_c^3} \frac{d^3\rho}{a_c^3}
\,\,e^{i\textbf{G}_m\cdot \boldsymbol{\rho}}
\,\,e^{-i\textbf{G}_{m'}\cdot \boldsymbol{\rho}'}
\,\,\,\,\,\,\,\,\,\,\,\,\,\,\,\,\,\,\,\,\,
\\
\frac{1}{N}\sum_{\ell=1}^N  \,
\frac{e^2}{|\vR_\ell+\boldsymbol{\rho}'-\boldsymbol{\rho} |}
e^{-i\vK\cdot(\vR_\ell+\boldsymbol{\rho}'-\boldsymbol{\rho})}\nn
\eea

\noindent $\bullet$ To calculate the above quantity, we first set $\vR_\ell+\boldsymbol{\rho}'=\vr'$. By noting that $e^{-i\textbf{G}_{m'}\cdot\boldsymbol{\rho}'}=e^{-i\textbf{G}_{m'}\cdot \vr'}$ since $e^{-i\textbf{G}_{m'}\cdot\vR_\ell}=1$ whatever  $\vR_\ell$, we  get
\bea 
\frac{1}{N}\sum_{\ell=1}^N\int_{a_c^3}\frac{d^3\rho'}{a_c^3}\frac{e^2}{|\vR_\ell+\boldsymbol{\rho}'-\boldsymbol{\rho} |}
e^{-i  (\vK+\textbf{G}_{m'}) \cdot (\vR_\ell+\boldsymbol{\rho}')}\,\,\,\,\,\,\,\,\,\,\,\,\,\,\,\,\,\,
\\
=\int_{L^3} \frac{d^3r'}{L^3}\frac{e^2}{|\vr'-\boldsymbol{\rho} |}e^{-i(\vK+\textbf{G}_{m'}) \cdot  \vr'}=e^{-i(\vK+\textbf{G}_{m'}) \cdot  \boldsymbol{\rho}}\,\,v_{\vK+\textbf{G}_{m'}}
\nn
\eea
with the Coulomb part  defined as
\be\label{41'}
v_{\vq}=\int_{L^3} \frac{d^3r}{L^3}e^{-i \vq \cdot \vr}\frac{e^2}{|\vr|}
\ee
The RHS of the above equation shows that $v_{\vq}$ scales as $e^2/L$ for $\vq=\v0$, while for nonzero $\vq$, its value, obtained by using the mathematical identity
\be\label{cheh_ex:22}
\frac{e^2}{|\vr|}=\frac{1}{(2\pi)^3}\int_{\infty}d^3q\frac{4\pi e^2}{q^2}e^{i\vq\cdot\vr}
\ee
is equal in the large sample limit, to
\be\label{42'}
v_{\vq\neq\v0}=\frac{4\pi e^2}{L^3 q^2}
\ee

When inserted into Eq.~(\ref{cheh_ex:17}), we end with
\be\label{cheh_ex:19}
W_\vK(\textbf{G}_{m'},\textbf{G}_m)=
 v_{\vK+\textbf{G}_{m'}}     \Big( \frac{1}{N}\sum_{\ell=1}^N \Big)
\int_{a_c^3} \frac{d^3\rho}{a_c^3} e^{i\boldsymbol{\rho}\cdot(\textbf{G}_m-\textbf{G}_{m'})}
\nn
\ee
\be\label{cheh_ex:19'}
=v_{\vK+\textbf{G}_{m'}}   
     \int_{L^3} \frac{d^3r}{L^3}e^{-i\vr\cdot(\textbf{G}_m-\textbf{G}_{m'})}=\delta_{m',m}\,\,v_{\vK+\textbf{G}_m} 
     \ee
%
%
%
%
 
 \noindent $\bullet$ 
 All this allows us to write the interband Coulomb scattering $V_{\vK}(\vk',\vk)$ as in Eq.~(\ref{17'}), with $V_{\vK,\textbf{G}_m}(\vk',\vk)$ equal to 
\bea
\label{35'} 
V_{\vK,\textbf{G}_m}(\vk',\vk)=
v_{\vK+\textbf{G}_m}
\,w^\ast_{\vK,\vk';\textbf{G}_m}\, w_{\vK,\vk;\textbf{G}_m}
\eea
for $v_\vq$ given in Eq.~(\ref{42'})   and $w_{\vK,\vk;\textbf{G}_m}$ given in Eq.~(\ref{cheh_ex:25}).

\subsection{From valence electron to hole\label{sec3G}}   

 Before calculating the $V_\vK(\vk',\vk)$ scattering amplitude explicitly, let us first turn from valence and conduction electrons to electrons and holes, in order to demonstrate that the interband Coulomb processes only act on electron-hole pairs that are in a spin-singlet state. This important point follows from the sign change that appears when transforming the destruction operator for valence electron into the creation operator for hole.

In the case of valence states with a threefold cubic degeneracy labeled by $\mu=(x,y,z)$ along the crystal axes, it is possible to show\cite{Monicbook,SYsec2021,Fetter} that the destruction of a valence electron with spatial index $\mu$, spin $s$ and wave vector $\vk$, corresponds to the creation of a $(\mu,-s,-\vk)$ hole within a sign change, namely
\be\label{cheh_ex:14_1}
\hat{a}_{\mu,v,\vk,s}=(-1)^{\frac{1}{2}-s}\,\hat{b}^\dag_{\mu,-\vk,-s} 
\ee    

When used in the sum over spins of Eq.~(\ref{cheh_ex:12}), that we rewrite as
\bea
\sum_{s',s}\hat{a}^\dag_{c,\vk',s'} 
\hat{a}^\dag_{v,\vk-\vK,s}  \hat{a}_{c,\vk,s} \hat{a}_{v,\vk'-\vK,s'}\nn\hspace{2cm}\\
=-\delta_{\vk',\vk}\sum_s \hat{a}^\dag_{c,\vk,s} \hat{a}_{c,\vk,s}\hspace{2.8cm}\label{cheh_ex:14_2}\\
+\sum_{s'} \hat{a}^\dag_{c,\vk',s'}\hat{a}_{v,\vk'-\vK,s'}\sum_s \hat{a}^\dag_{v,\vk-\vK,s} \hat{a}_{c,\vk,s} \nn
\eea 
we see appearing the creation operator for an electron-hole pair having a center-of-mass wave vector $\vK$ and an electron wave vector $\vk'$, in a spin-singlet state $(S=0,S_z=0)$, namely
\bea
\sum_{s'} \hat{a}^\dag_{c,\vk',s'}\hat{a}_{v,\vk'-\vK,s'}\!\!&=&\!\!\hat{a}^\dag_{\vk',\frac{1}{2}}\hat{b}^\dag_{\vK-\vk',-\frac{1}{2}}-\hat{a}^\dag_{\vk',-\frac{1}{2}}\hat{b}^\dag_{\vK-\vk',\frac{1}{2}}\nn\\
&\equiv& \sqrt{2}\,\,\hat{B}^\dag_{\vK,\vk';S=0,S_z=0}\label{cheh_ex:14_3}
\eea

As a result, the interband Coulomb interaction given in  Eq.~(\ref{cheh_ex:12}) ultimately reads in terms of electron-hole pairs as 
\bea\label{cheh_ex:14_4}
\widehat{V}^{(inter)}_{e-e}&=& -  \sum_\vK  \sum_{s\vk} V_\vK(\vk,\vk)  \hat{a}^\dag_{c,\vk,s} \hat{a}_{c,\vk,s}
\nn\\
&+&  
2\sum_\vK \sum_{\vk',\vk} V_\vK(\vk',\vk) \hat{B}^\dag_{\vK,\vk';0,0} \hat{B}_{\vK,\vk,0,0}
\eea
The first term brings a contribution to the kinetic energy of the conduction electrons that accounts for the Coulomb interaction of a conduction electron with all valence electrons, contribution that is forgotten when speaking in terms of holes. The second term demonstrates that the interband Coulomb interaction brings a \textit{repulsive} interaction between electron-hole pairs that are in the spin-singlet state $(S=0, S_z=0)$: The ``incoming'' pair is made with a $\vk$ electron and the ``outgoing'' pair with a $\vk'$ electron, both pairs having the same center-of-mass wave vector $\vK$, which also is the wave vector transfer of the interband scattering at hand, as seen from Fig.~\ref{fig_1}.

\section{Calculation of the interband Coulomb scattering\label{sec3}}

\subsection{In the $\textbf{G}_m$ reciprocal space }

The $\textbf{G}_m$ space is the appropriate space to calculate the interband Coulomb scattering amplitude $V_{\vK}(\vk',\vk)$ in the small $\vK$ limit because $\vK+\textbf{G}_m$ in Eq.~(\ref{35'}) never cancels in this limit: indeed, the ($2\pi/a_c$) scale for $\textbf{G}_m$ vectors is very large compared with the ($2\pi/L$) scale for $\vK$ vectors. So, $v_{\vK+\textbf{G}_m}$ defined in Eq.~(\ref{41'}) is unambiguously equal to $4\pi e^2/L^3 |\vK+\textbf{G}_m|^2$ when $\vK \rightarrow \bf0$. It then becomes clear from Eq. (\ref{35'}) that if a singular behavior has to exist for $V_{\vK \rightarrow \bf0}(\vk',\vk)$, it has to come from the $\textbf{G}_0=\v0$ term of the $\textbf{G}_m$ sum in Eq.~(\ref{17'}), the remaining part of the sum going smoothly to its  $\vK=\v0$ value given by
\bea\label{cheh_ex:26_1}
  \sum_{m\not=0}V_{\vK\rightarrow {\bf0},\textbf{G}_m}(\vk',\vk)=\hspace{5cm}
\\ 
\sum_{m\not=0} \frac{4\pi e^2}{L^3G_m^2}
\lan u_{c,\vk'}|e^{i\textbf{G}_m\cdot\vr}|u_{v,\vk'}\ran\lan u_{v,\vk}|e^{-i\textbf{G}_m\cdot\vr}|u_{c,\vk}\ran
\nn 
\eea

By contrast,  $V_{\vK,\textbf{G}_0=\v0}(\vk',\vk)$ is indeed singular in the small $\vK$ limit: it depends on the $\vK$ direction with respect to the crystal axes. To show it, we first note that\cite{Monicbook}
\be\label{cheh_ex:26_3}
|u_{n,\vk+\vK}\ran\simeq |u_{n,\vk}\ran+\frac{\hbar}{m_0}\sum_{n'\not=n}\frac{|u_{n',\vk}\ran \lan u_{n',\vk}|\vK\cdot\hat{\vp}  |u_{n,\vk}\ran}{\va_{n,\vk}-\va_{n',\vk}}
\ee
as obtained from first-order perturbation theory in the $\vk\cdot\vp $ formalism. So, from
\be
 \label{33'}
 V_{\vK,\textbf{G}_0=\bf0} (\vk',\vk)=
\frac{4\pi e^2}{L^3K^2}\lan u_{c,\vk'}|u_{v,\vk'-\vK}\ran\lan u_{v,\vk-\vK}|u_{c,\vk}\ran
\ee
as obtained from Eqs.~(\ref{cheh_ex:25},\ref{35'}), we get
 \bea
 \label{33}
 V_{\vK\rightarrow \bf0,\textbf{G}_0=\bf0} (\vk',\vk)\simeq\hspace{5cm}
\\
 \frac{4\pi e^2}{L^3K^2} \lan u_{c,\vk'}|\frac{ \hbar \vK\cdot\hat{\vp}}{m_0E_{gap}}|u_{v,\vk'}\ran\lan u_{v,\vk}|\frac{ \hbar  \vK\cdot\hat{\vp}}{m_0E_{gap}}|u_{c,\vk}\ran
\nn
\eea
since $\va_{c,\vk}-\va_{v,\vk}$ is close to the band gap $E_{gap}$. 

To go further, we note that, for small $\vk$'s, which are the relevant wave vectors out of which Wannier excitons are constructed, the vector $\lan u_{v,\vk}|\hat{\vp}|u_{c,\vk}\ran$ differs from zero due to parity; so, for cubic GaAs-like semiconductors having a threefold valence level with states labeled as $\mu=(x,y,z)$ along the crystal axes, that transform like $\{x,y,z\}$ in the crystal symmetry operations, we  have
\be\label{33_1}
\lan u_{v,\mu,\v0}|\vK\cdot\hat{\vp}|u_{c,\v0}\ran=K_{\mu}\lan u_{v,\mu,\v0}|\hat{p}_{\mu}|u_{c,\v0}\ran \equiv K_{\mu}P_{v,c}
\ee
the $K_{\mu}$ prefactor being $\mu$ independent due to cyclic symmetry in a cube. 

Inserting the above result into Eq.~(\ref{33}) yields, for $(\vk',\vk)$ small,
\bea
 \label{35}
  V_{\vK\rightarrow {\bf0},\textbf{G}_0=\bf0} (\mu',\vk';\mu,\vk)
\simeq \frac{4\pi e^2}{L^3}
 \, \, \Big|\frac{\hbar  P_{c,v}}{m_0 E_{gap}}\Big|^2
 \;\frac{K_{\mu'} K_\mu}{K^2}
\eea 
So, for $\vK\rightarrow \bf0$, the $\textbf{G}_0=\bf0$ term in the $\textbf{G}_m$ expansion of the interband Coulomb scattering, depends on the $\vK$ direction with respect to the crystal axes: This proves that the interband Coulomb scattering indeed is highly singular in this limit.

\subsection{In the $\vR_\ell $ real space}

Let us now go back to the interband Coulomb scattering given in Eq.~(\ref{cheh_ex:14}) and show how this singular behavior also appears in the $\vR_\ell $ expansion of the interband Coulomb scattering amplitude $V_\vK(\vk',\vk)$ when written in terms of $V_\vK(\vk',\vk; \vR_\ell)$.

\subsubsection{Intracell $\vR_\ell= \v0$ contribution}

When $\vK$ goes to zero, the contribution to $V_\vK(\vk',\vk)$ that comes from the $\vR_\ell= \v0$ term, goes smoothly to its $\vK=\v0$  value that reads
 \bea\label{cheh_ex:32}
V_{\vK\rightarrow \bf0}(\vk',\vk;\vR_\ell= \v0) =    \hspace{4cm}\\
  \frac{1}{N}\iint_{a_c^3}\frac{d^3\rho'}{a_c^3} \frac{d^3\rho}{a_c^3} \frac{e^2}{|\boldsymbol{\rho}'-\boldsymbol{\rho} |}w^\ast_{\v0,\vk'}(\boldsymbol{\rho}')w_{\v0,\vk}(\boldsymbol{\rho})
  \nn  
\eea
It corresponds to the Coulomb interaction between two charge distributions inside a unit cell. 

\subsubsection{Intercell  $\vR_\ell\neq\v0$ contribution}

\noindent $\bullet$ The sum of all $\vR_\ell\not=\bf0$ terms in the $V_\vK(\vk',\vk )$ expansion corresponds to contributions from the rest of the lattice. Using Eq.~(\ref{17}), it reads 
 \bea
V^{(latt)}_\vK(\vk',\vk )=\frac{1}{N}\iint_{a_c^3}\frac{d^3\rho}{a_c^3} \frac{d^3\rho'}
{a_c^3}  
w^\ast_{\vK,\vk'}(\boldsymbol{\rho}')
w_{\vK,\vk}(\boldsymbol{\rho})
\nn
\\
e^{-i\vK\cdot(\boldsymbol{\rho}'-\boldsymbol{\rho})}
\sum_{\vR_\ell\not=\bf0} \frac{e^2}{|\vR_\ell+\boldsymbol{\rho}'-\boldsymbol{\rho} |}e^{-i\vK\cdot\vR_\ell}\label{cheh_ex:33}
\hspace{0.2cm}
\eea

\noindent $\bullet$ To calculate the $\vR_\ell$ sum, we first note that any nonzero $|\vR_\ell|$ is large compared to $|\boldsymbol{\rho}-\boldsymbol{\rho}'|$ since $\boldsymbol{\rho}$ and $\boldsymbol{\rho}'$ are restricted to a cell. So, we can expand $1/|\vR_\ell+\boldsymbol{\rho}-\boldsymbol{\rho}' |$ in power of $1/R_\ell$. This expansion follows from
\bea
\label{cheh_ex:36}
\frac{1}{|\vR_\ell+\boldsymbol{\rho}|}
=
\frac{1}{\sqrt{R_\ell^2+2 \vR_\ell \cdot \boldsymbol{\rho}+\rho^2}}\hspace{3.3cm}
\\
=
\frac{1}{R_\ell}\left[1-\frac{1}{2}\frac{2 \vR_\ell\cdot\boldsymbol{\rho}{+}\rho^2}{R_\ell^2 }+\frac{3}{8}\left(\frac{2 \vR_\ell\cdot\boldsymbol{\rho}}{R_\ell^2}\right)^2+\cdots\right]\nn
\eea

\noindent $\bullet$ When used into Eq.~(\ref{cheh_ex:33}), we see that the dominant terms of the resulting $1/R_\ell$ expansion come from terms in $\boldsymbol{\rho}'\,\boldsymbol{\rho}$ provided that 
\be\label{cheh_ex:37_3}
\int_{a_c^3}\frac{d^3\rho}{a_c^3} \,\boldsymbol{\rho}\, w_{\v0,\vk}(\boldsymbol{\rho})
\ee
differs from zero; indeed, all the other terms of the $1/R_\ell$ expansion give zero when integrated over $(\boldsymbol{\rho}',\boldsymbol{\rho})$ in the $\vK\rightarrow \bf0$ limit. The two dominant terms in $\boldsymbol{\rho}\,\boldsymbol{\rho}'$ are
\be\label{cheh_ex:37_1}
\frac{1}{R_\ell^3}\left[\boldsymbol{\rho}'\cdot\boldsymbol{\rho}-3\,\,  \left( \boldsymbol{\rho}'   \cdot \frac{\vR_\ell}{R_\ell}  \right)  \,\,     \left( \boldsymbol{\rho}\cdot\frac{\vR_\ell}{R_\ell}   \right)  \right]
\ee

To understand why the integral (\ref{cheh_ex:37_3}) differs from zero for GaAs-like semiconductors, we must remember that these materials have a threefold valence level, its states labeled as $(\mu,\vk)$ having a $\mu$ spatial symmetry. Equation (\ref{cheh_ex:37_3}) then appears, for \textbf{k} small, as
\be
\label{cheh_ex:37_4}
|e|\int_{a_c^3}\frac{d^3\rho}{a_c^3}  \boldsymbol{\rho} \,u^\ast_{\mu,v,\v0}(\boldsymbol{\rho})u_{c,\v0}(\boldsymbol{\rho})
= d_{ v,c}  \textbf{ e}_{\mu}
\ee
the $d_{ v,c}$ prefactor being $\mu$ independent due to cyclic symmetry in a cube. This prefactor physically corresponds to the dipole moment of the valence-conduction distribution.

\noindent $\bullet$ When used into Eq.~(\ref{cheh_ex:33}), this gives the interband Coulomb scattering amplitude in the $\vK\rightarrow \bf0$ limit, for ($\vk',\vk$) small, as
\be\label{cheh_ex:37_5}
V^{(latt)}_{\vK\rightarrow \bf0}(\mu',\vk';\mu,\vk )\simeq\frac{ |d_{ v,c}|^2}{N}S_{\vK\rightarrow \bf0} (\mu',\mu)
\ee
with $S_\vK (\mu',\mu)$ given by
\be\label{cheh_ex:37_6}
S_\vK (\mu',\mu)=\!\!\sum_{\vR_\ell\not=\bf0}\!\!\frac{e^{-i\vK\cdot\vR_\ell}}{R^3_\ell}\left[\delta_{\mu',\mu}-3
\left(\textbf{e}_{\mu'} \cdot
\frac{\vR_\ell}{R_\ell}\right)\!\!
\left(\textbf{e}_{\mu} \cdot
\frac{\vR_\ell}{R_\ell}\right) \right]
\ee
This sum is highly singular in the small $\vK$, as seen from 
\begin{subeqnarray}\label{cheh_ex:47}
S_{\vK=\bf0}(\mu',\mu)&=& 0\slabel{cheh_ex:47_0}\\
S_{\vK\rightarrow{\bf0}}(\mu',\mu)&=& \frac{4\pi}{3a_c^3}\left(3\frac{K_{\mu'}K_{\mu}}{K^2}-\delta_{\mu',\mu}\right)\slabel{cheh_ex:47_2}
\end{subeqnarray}
which is explicitly derived in the Appendix \ref{app:C}.

\subsubsection{Link with the calculation in $\textbf{G}_m$ space}

The above result shows that in the small $\vK$ limit, the long-range Coulomb contribution calculated from the sum of all $\vR_\ell\not=\bf0$ terms, has the same $K_{\mu'}K_{\mu}/K^2$ singular behavior as $V_{\vK\rightarrow \bf0,\v0}(\mu',\vk';\mu,\vk)$ calculated in the $\textbf{G}_m$ reciprocal space, as given in Eq.~(\ref{35}). The singular behavior comes from the large $ \vR_\ell$ terms in the sum over lattice vectors while in $\textbf{G}_m$ space, it appears through the unique $\textbf{G}_0=\v0$ term.
 
 The prefactors of the $K_{\mu'}K_{\mu}/K^2$ singularity in the $\textbf{G}_0=\v0$ term and in the $\vR_\ell\neq\v0$ sum also are equal: Indeed, according to Eq.~(\ref{35}) and Eqs.~(\ref{cheh_ex:37_5},\ref{cheh_ex:47}), we do have
\be
 \label{59}
 \frac{4\pi e^2}{L^3}
 \left|\frac{\hbar  P_{c,v}}{m_0 E_{gap}}\right|^2
=\frac{ \left|d_{c,v}\right|^2}{N} \, \,   \frac{4\pi}{a_c^3}
\ee
This equality follows, in a nontrivial way, from
\be
 \label{59_1}
 E_{gap}=i\frac{\hbar  P_{c,v}}{m_0 d_{c,v}/|e|}
 \ee 
 as shown in Appendix \ref{app:D}.

Using the above result, we can also show that the $\vK\rightarrow \bf0$ limit of $\textbf{G}_m\not=\bf0$ sum in Eq.~(\ref{cheh_ex:26_1}) is equal to the limit of the intracell $\vR_\ell=\bf0$ term in Eq.~(\ref{cheh_ex:32}). This establishes that the interband Coulomb scattering can indeed be calculated within the two approaches, either short-range plus long-range interactions in the real space, or better with respect to the singularity,  $\textbf{G}_m=\bf0$ plus $\textbf{G}_m\not=\bf0$ interactions in the reciprocal space.

\section{Past and future \label{sec5}}

\subsection{State of the art}

The singularity of the interband Coulomb scattering in the small wave-vector transfer limit, was first pinned down in the study of excitons. Early seminal works\cite{Wannier1937,Elliot1961,Rashba1959,Moskalenko1959} on excitons were devoted to finding a microscopic formalism that starting from a fully occupied semiconductor valence band, would support the reduction of its lowest excited states, many-body in essence, to a two-body state, one electron and one hole coupled by a na\"{i}ve Coulomb interaction, from which hydrogen-like bound states follow naturally. In the 60's, the common way to describe a many-body system like $N$ valence electrons having one electron excited in the conduction band, was through Slater determinants in the first quantization formalism. Its far simpler description through particle operators in second quantization was not yet popular. One great advantage of the operator formalism is to avoid the very many minus signs associated with ``exchange'', that appear when calculating matrix elements between Slater determinants, these minus signs being automatically taken care of by the anticommutation relation between fermion operators.

In these seminal works, the Coulomb interaction given in Eq.~(\ref{cheh_ex:2}) was handled within the first quantization formalism, by calculating its matrix elements between Slater determinants for  $(N-1)$ electrons in a valence state and one electron in a conduction state. One then sees that the Coulomb interaction generates ``direct'' terms in which a scattered valence electron  stays valence electron,  and a scattered conduction electron stays conduction electron. These direct terms correspond to the \textit{intraband} Coulomb scatterings that produce the Wannier exciton binding.

One also sees that due to the determinant form of the many-fermion wave function, an electron which is in a valence state on the RHS of the matrix element, can  appear in a conduction state on the LHS --- with a minus sign due to the wave function antisymmetry: This change from valence to conduction state, corresponds to an \textit{interband} Coulomb process, that is, an \textit{electron} exchange between the two bands. This electron exchange is definitely not an electron-hole exchange because an electron and a hole, which is a valence electron absence in a full valence band, are not identical fermions to possibly suffer a quantum exchange.  A possible reason for this physically incorrect denomination is that the hole concept is enforced into the electron Slater determinants  in a way that is both jarring and insecure\cite{Knoxbook}.  Slater determinants have been proposed as a wise way to represent the antisymmetric wave function of indistinguishable fermions, not the wave function of different fermions like electrons and holes. As a result, instead of an electron exchange between valence and conduction bands, as it is, this term has been seen as an electron-hole exchange, that it is not. In that respect, the Feynman diagram shown in Fig.~\ref{fig_8}, that represents a Coulomb scattering between conduction and valence bands, evidences that such an interband scattering does correspond to an \textit{electron-hole pair} exchange, one pair being replaced by another pair.

The transformation of a valence-electron absence into a hole is tricky: One starts with $(N-1)$ fermions but ends with one fermion only. Even if  physical intuition leads us to accept that the hole must have a positive charge and a positive mass, this transformation goes along with other sign changes that are not intuitive or even ignored, such as the sign change (\ref{cheh_ex:14_1}) between valence electron and valence hole operators that directly leads to interband Coulomb processes occurring between spin-singlet electron-hole pairs only. In our opinion, the only clean and secure way to speak in terms of holes, is through the operator formalism of the second quantization.

Still, it is possible, within the Slater determinants of the first quantization, to identify Coulomb processes that only exist for  electron-hole pairs in the spin-singlet configuration. It is moreover possible to write down the analytical expression of the associated scattering, and through its calculation, to show that this term is singular when the wave vector $\textbf{K}$ of the scattered electron-hole pair goes to zero, this pair wave vector also being the wave vector transfer of the associated Coulomb process, as evidenced in Fig.~\ref{fig_8}. The calculation of this singularity has been mostly performed in terms of Bloch wave functions with periodicity written via  $\textbf{R}_\ell$ vectors in the real space. Yet, the proper way to handle  periodic functions is by turning to the $\textbf{G}_m$ vectors of the reciprocal space, as we do in this paper. The singularity then comes from the unique $\textbf{G}_0=\textbf{0}$ term. When performed in the real space, a far heavier calculation gives the interband term  as a sum over all $\textbf{R}_\ell$ lattice vectors. The large $\textbf{R}_\ell$'s, associated with ``long-range'' Coulomb processes, are responsible for the singularity in the small $\textbf{K}$ limit, while the $\textbf{R}_\ell=\textbf{0}$ term, associated with ``short-range'' processes, is regular in this limit. The commonly quoted result for the small-$\textbf{K}$ singularity is correct. Yet, as we have not been able to find a precise derivation of this mathematically nontrivial result in the literature, we  propose a  derivation in the Appendix \ref{app:C}.

The consequences of the ``electron-hole  exchange'' interaction on Wannier excitons were initiated by Moskalenko and Tolpygo\cite{Moskalenko1959}, who introduced this interaction as a correction to the intraband Coulomb processes responsible for the exciton binding. They correctly showed that while the intraband Coulomb processes exist for whatever the carrier spins, the interband Coulomb processes exist for  excitons made of spin-singlet electron-hole pairs only. These pioneering works were soon followed by more studies\cite{Elliot1961,Onodera1967,Dos1968,Skettrup1970,Pikus1971,Rohner,Denisov1973,Makarov1968,Cho1976}, in particular Onodera and Toyozawa\cite{Onodera1967},  Pikus and Bir\cite{Pikus1971} and Denisov and Makarov\cite{Denisov1973}. These authors formulated the ``electron-hole exchange'' interaction in terms of Bloch states; they divided its expression in the $\vR_\ell$  space\cite{Onodera1967,Skettrup1970} through ``long-range'' and ``short-range'' terms, and also in the $\textbf{G}_m$ reciprocal space\cite{Pikus1971,Denisov1973}, through the $\textbf{G}_0=\textbf{0}$ term and a sum of $\textbf{G}_m\neq \textbf{0}$ terms, as we here do. All these works are written within the first quantization formalism, except Ref.\onlinecite{Denisov1973} that uses the second quantization. In this work, the authors attributed the exciton longitudinal-transverse splitting  to the nonanalytical part of the Coulomb exchange terms; however, the spin-dependent phase factor appearing in the relation (\ref{cheh_ex:14_1}) between electron and hole states was not given and the distinction between singlet and triplet exciton was not explicitly stated. More recent works\cite{Hanamura,Haug&Schmitt-Rink,Haug&Koch} also discuss excitons within the second quantization formalism; these works mentioned, but did not discuss the interband Coulomb terms; they actually focused on semiconductor optical properties; the exciton problem was approached only through the interband polarization induced by light in the semiconductor, so that only the bright excitons were concerned, by construction.

We wish to stress that these previous works did not mention the important role played by the one-body average electron-electron Coulomb interaction $\overline{V}_{e-e}$ in Eq.~(\ref{9'}), that has to be introduced to define the periodic Bloch-state basis. One reason can be that all these works skip the very first part of the present work on the Coulomb interaction for semiconductor electrons and the crucial fact that what is called ``Coulomb interaction'' $\mathcal{V}_{Coul}$ in a semiconductor is \textit{not} the two-body Coulomb interaction (\ref{cheh_ex:2}), but its difference with the average Coulomb interaction, as defined in Eq.~(\ref{9}). The introduction of this average electron-electron interaction allows us to cleanly eliminate processes associated with zero-wave-vector transfers, for both the intraband and interband Coulomb scatterings, as  here shown. These spurious divergent $(\textbf{q}=\textbf{0})$ terms, which cancel out exactly thanks to the average Coulomb interaction, appear in previous calculations, Following G. D. Mahan \cite{Mahan}, the authors of Ref.\onlinecite{Haug&Koch} eliminated these spurious terms in the particular case of a jellium model in which the ions are considered as forming a uniform positive charge background, thus neglecting the discrete lattice structure of ions in a crystal.

To conclude this short ``state of the art'', we also wish to mention the longstanding but still much debated question of the Coulomb screening for long-range and short-range processes, induced by the electrons that remain in the valence band\cite{Knoxbook,Dos1968,Skettrup1970,Pikus1971,Rohner,Denisov1973,Makarov1968,Kohn1958,Abe1962,Sham1966}. As this screening involves the excitations of the whole valence band through virtual electron-hole pairs, its impact on both, the intraband scatterings and the interband scatterings, can only be properly derived within a full many-body approach through the second quantization formalism. Up to now, the problem of the Coulomb screening by valence electrons in the presence of a bound-state exciton has been solved by inserting the bubble processes of the random-phase-approximation (RPA), dominant in the dense limit only, inside the intraband ladder processes that are dominant in the dilute limit of one electron and one hole. It is clear that this proposed solution  is  highly questionable  because it selects processes that are dominant in two opposite limits. The important but very difficult problem of what should be done in the case of long-range and short-range interband processes and the na\"{\i}ve idea that no screening should occur inside a unit cell, will be addressed in a further work devoted to Coulomb screening in a semiconductor.

\subsection{Interband Coulomb interaction in the exciton physics}

\noindent $\bullet$ To study the effects of the interband Coulomb interaction on excitons in a realistic way, it is necessary to take into account the spin-orbit interaction. This interaction splits the $(3\times2)$-fold subspace of $(\mu,s)$ holes, into a fourfold subspace and a twofold subspace\cite{Cardona}. In the spherical approximation that neglects the warping\cite{DresselhausPR1955,Dresselhaus1956,Luttinger,Baldereschi_prl,Baldereschi_prb}, the fourfold subspace is made of heavy and light holes that are associated with a hole index $\mathcal{J}=\pm3/2$ and $\mathcal{J}=\pm1/2$ quantized along the hole wave vector, the corresponding eigenstates being a mixture of ($\mu,s$) holes. Obviously, this mixing is going to hide the fact that the interband Coulomb interaction only acts on spin-singlet electron-hole pairs.

\noindent $\bullet$ Moreover, electron-hole pairs with the same center-of mass wave vector $\vK$ but different hole wave vectors $\vk_h$, interact through intraband Coulomb processes, to ultimately form excitons having a center-of-mass wave vector equal to $\vK$. The thorny problem is that the intraband Coulomb scatterings, diagonal between $\mu$ holes, do not stay diagonal between 
 heavy and light holes\cite{SYprb2021}: Indeed, a heavy hole can turn light under a Coulomb scattering. This mass change prevents using the hydrogen-like procedure to derive the exciton eigenstates because, unless the two holes are taken with the same mass, it is not possible to isolate a relative-motion wave vector for the electron-hole pair.

The effect of the hole mass difference on exciton having a finite center-of-mass wave vector $\vK$ is still an open problem\cite{Shiauprbl2022}. It is necessary to first solve this major intraband Coulomb problem before tackling the consequences of the interband Coulomb singularity on free pairs, out of which the excitons are formed.

\noindent $\bullet$ The exciton physics actually splits into two regimes in which  the consequences of the interband Coulomb singularity deserves to be reconsidered\cite{Ulbrich1977} for two different reasons.

\noindent \textbf{\textit{(1)}} If the lifetime of the exciton state with  wave vector $\vK$  is long compared to the time associated with the exciton recombination into a photon, a mixed state  called polariton is formed out of a photon and a bright exciton\cite{Andreani1988}. To derive this mixed state, the exciton-photon coupling has to be treated exactly, \textit{i.e.}, not as lowest order in perturbation.  The intraband Coulomb interaction enters the polariton regime through the formation of exciton out of free electron-hole pairs. The interband Coulomb interaction also has to enter into play. 

To catch it, we may note that the photon is coupled to bright exciton having a center-of-mass wave vector $\vK$ equal to the photon wave vector $\textbf{Q}_{ph}$. For $\vK$ along the crystal axis $\textbf{e}_z$, that is, for $K_z=K$ and $K_x=K_y=0$, the interband Coulomb interaction shifts the bright exciton having a $\mu=z$ hole, but does not affect the excitons having a $x$ or $y$ hole, as seen from the $K_{\mu'}K_\mu$ factor in the Coulomb scattering of Eq.~(\ref{35}). This $K_{\mu'}K_\mu$ factor thus produces an energy increase to ``longitudinal'' exciton with hole index $z$ along $\vK$ compared to ``transverse'' exciton with hole index $(x,y)$ orthogonal to $\vK$. We must then remember that, due to spatial symmetry conservation, a photon with wave vector $\textbf{Q}_{ph}$ along $\textbf{e}_z$ is not coupled to  longitudinal excitons having a $z$ hole, because such a photon has a polarization vector in the $(x,y)$ plane. As a result, the polariton mode for photon $\textbf{Q}_{ph}=\vK$ along $\textbf{e}_z$ is made of transverse bright excitons only, the longitudinal excitons with hole index $z$ along $\vK$ being unaffected by the presence of these photons.

Then comes the fact that the standard way to derive the polariton eigenstates is to use an approximate Hamiltonian, in which the electron-photon interaction is reduced to its \textit{resonant} linear coupling to the ground-exciton level. It is clear that this reduction can only be valid in the vicinity of the exciton-photon resonance. To extend the validity of the polariton energy obtained from this approximate Hamiltonian, from \textbf{K}=\textbf{0} to \textbf{K} infinite, that is, far from resonance, is highly questionable. Its extension is even more problematic because the linear electron-photon coupling diverges in the $\vK\rightarrow \bf0$ limit\cite{Monicbook}. As a direct consequence, the energy of the exciton branch would then go to infinity. To get rid of this unphysical behavior properly, it is necessary to include not only \textit{all} the exciton levels but also the quadratic coupling  as well as all \textit{nonresonant} electron-photon processes. Their inclusion is dramatic because the nonresonant couplings, with one photon created along with the creation of a photon or an exciton, transform the one exciton-one photon problem into a many-body problem, for which the scenario of a single longitudinal exciton being related to the $\vK\rightarrow \bf0$ limit of the upper polariton branch, is  hard to accept.

Of course, we can choose to completely overlook this major many-body problem through replacing the semiconductor Hamiltonian by an effective Hamiltonian for \textit{noninteracting bosonized} excitons having transverse and longitudinal energies. It is then possible to show that, under linear and quadratic, resonant \textit{and} nonresonant photon couplings to all exciton levels, the second polariton branch  goes in the small $\textbf{Q}_{ph}$ limit to the uncoupled longitudinal exciton, within a correction induced by the photon couplings to all the other exciton levels. This correction, of the order of the change of the lowest polariton branch compared to the bare photon energy, has not been mentioned in the previous works\cite{Quattropani1986,Bassani1986} dealing with this problem because these works incorrectly drop the other exciton levels from the very first line. So, the link between the polariton energies for small and large wave vectors, and the singularity of the bright pair scattering in the small \textbf{K} limit, responsible for their longitudinal-transverse splitting, only is approximate.

\noindent \textbf{\textit{(2)}} The exciton regime in which photons are absorbed to form excitons is not any simpler. Although missed for a long time, the exciton-photon interaction also plays a role in this regime. Indeed, two exciton-photon interactions produce an interband process similar to the interband Coulomb process\cite{MoniqueEPL2022}. As a result, they participate in the energy shift suffered by bright electron-hole pairs. They even play a crucial role as they completely wash out the singular behavior of the interband Coulomb scattering with respect to the exciton wave vector direction. Since dark pairs do not suffer interband scatterings, the interband Coulomb interaction, along with the interband electron-photon interaction, produces the observed splitting between bright and dark excitons. As a result, this bright-dark splitting is not directly related to the so-called ``electron-hole exchange''. 

\noindent $\bullet$  All this shows that there is a long way from free electron-hole pairs with holes labeled along the crystal axes, as considered here, to excitons made of heavy and light holes that result from the spin-orbit interaction, with nondiagonal \textit{intra}band Coulomb processes between heavy and light holes\cite{SYprb2021}. This  interesting but quite complex extension of the present work on interband Coulomb interaction to the exciton physics, deserves further investigation.

The impact of the interband Coulomb interaction on the exciton physics extends more broadly to semiconductor quantum wells. In direct type-I quantum wells, the longitudinal-transverse exciton splitting depends linearly\cite{Sham1993} on the exciton wave vector $\vK$, due to the $1/q$ dependence of the Coulomb interaction in 2D systems\cite{Monicbook}. Its orientational dependence leads to the relaxation between circularly polarized exciton states, analogous to the Dyakonov-Perel spin relaxation of the conduction electrons in quantum wells\cite{Dyakonov1986}. Recently, the interband Coulomb interaction has also become a subject of interest for two-dimensional semiconductor materials. In direct-gap monolayer transition metal dichalcogenides\cite{Glazov2014,Qiu2015,Echeverry2016}, the relevant conduction and valence electrons close to the gap belong to two different (K,K$^\prime$)  valleys at the edge of the Brillouin zone that correspond to each other by time reversal symmetry. These two valleys and the strong spin-orbit interaction result in exciton states with a rich spin-valley texture\cite{Glazov2014,Qiu2015,Echeverry2016}. Spin-singlet excitons that are made of a conduction electron and a missing valence electron in the same K or K$^\prime$ valley, termed direct excitons, can be coupled by the interband Coulomb interaction\cite{Glazov2014,Qiu2015}, to produce a linear-$\vK$ longitudinal-transverse splitting. However, for spin-singlet indirect excitons that are made of a conduction electron and a missing valence electron from different valleys\cite{Pengkeprb2022}, the $\textbf{G}_0=\bf0$ term in the interband Coulomb interaction does not have any singularity.

\section{Conclusion}

In the first part of this work, we provide an \textit{ab initio} derivation of the interband Coulomb scattering between conduction electron and valence hole characterized by a spin $s$ and a spatial index $\mu=(x,y,z)$ along the cubic crystal axes of GaAs-like semiconductors. This interband scattering physically corresponds to exchanging two electron-hole \textit{pairs} with center-of-mass wave vector equal to the scattering wave-vector transfer. 

The quantum approach we here present, clarifies two important points related to the so-called ``electron-hole exchange''. 

\noindent (\textit{1}) By introducing the notion of hole through the second quantization formalism --- which is the clean way to do it --- we readily find that the interband Coulomb interaction only acts on electron-hole pairs that are in a spin-singlet state, in contrast to spin-triplet pairs that do not suffer interband processes. This mathematically follows from the sign change that appears in Eq.~(\ref{cheh_ex:14_1}) when turning from valence-electron destruction operator to hole creation operator. 

\noindent (\textit{2})  By comparing the calculations of the interband Coulomb scattering written in terms of lattice vectors $\vR_\ell$ and in terms of reciprocal vectors $\textbf{G}_m$, we evidence that periodic systems are better handled in the reciprocal space than in the real space: indeed, the singular behavior of the interband Coulomb scattering that scales in the small electron-hole pair wave vector limit,  as
\be
\frac{K_{\mu'}K_{\mu}}{K^2} 
\ee
for a $\vK$  pair that starts with a $\mu$ hole and ends with a $\mu'$ hole, has been established through a ``long-range'' sum over all lattice vectors;  in the reciprocal space, this singularity only comes from the $\textbf{G}_0=\bf0$ term,  far simpler to calculate than the tricky $\vR_\ell\neq \bf0$ sum.

 In this work, we also question the well-accepted consequences of this singular interband scattering not only in the polariton regime through a possible link with the exciton longitudinal-transverse splitting, but also in the exciton regime through a possible link with the bright-dark exciton splitting. We show that some aspects of this singular exciton-photon coupling remain improperly addressed. In view of the growing importance of excitons in quantum information technology, the present work should stimulate more fundamental research on semiconductor excitons and its interaction with light.

\appendix

\section{Singular sum in Eq.~(\ref{cheh_ex:37_6}) \label{app:C}} 

We here show that, in the small $\vK$ limit, the sum defined in Eq.~(\ref{cheh_ex:37_6}) has the singular behavior given in Eq.~(\ref{cheh_ex:47}).

To do it, we consider an arbitrary set of orthonormal vectors $({\bf x},{\bf y},{\bf z})$ and we take $\boldsymbol{\mu}$ and  $\boldsymbol{\mu}'$ as any of these three vectors. For $R_\mu=\vR\cdot \boldsymbol{\mu}$, the following sum over all lattice vectors $\vR$
\bea\label{app:1}
 S_\vK(\mu',\mu)&=&\sum_{\vR\not=\bf0}\frac{e^{-i\vK\cdot \vR}}{R^3}\left(\delta_{\mu',\mu}-3 \frac{R_{\mu'}R_\mu}{R^2}\right)\\
 &=&\sum_{\vR\not=\bf0}\frac{e^{-i\vK\cdot \vR}}{R^3}\left(\boldsymbol{\mu}'\cdot \boldsymbol{\mu}-3 \frac{(\vR\cdot \boldsymbol{\mu}')( \vR\cdot \boldsymbol{\mu})}{R^2}\right)\nn
 \eea
is singular  in the $\vK\rightarrow \bf0$ limit, that is, the value calculated for $\vK=\bf0$ differs from its limiting value for $\vK\rightarrow \bf0$.

\subsubsection{Calculation for $\vK=\bf0$}

Let us first show that, the $S_\vK(\mu',\mu)$ sum cancels for $\vK=\bf0$. To do it, we introduce the cubic crystal  axes $({\bf X}, {\bf Y}, {\bf Z})$ and expand $(\vR,\boldsymbol{\mu}',\boldsymbol{\mu})$ on these axes as
\bea
\vR&=&R_X {\bf X}+R_Y {\bf Y}+R_Z {\bf Z}\label{app:2}\\
\boldsymbol{\mu}&=& \mu_X {\bf X}+\mu_Y {\bf Y}+\mu_Z {\bf Z}\label{app:3}
\eea
with a similar result for $\boldsymbol{\mu}'$. This gives $\vR\cdot \boldsymbol{\mu}=R_X \mu_X+ R_Y\mu_Y+R_Z\mu_Z$. So, the product $(\vR\cdot \boldsymbol{\mu}' )(\vR\cdot \boldsymbol{\mu})$ in Eq.~(\ref{app:1}) reads 
\bea
(\vR\cdot \boldsymbol{\mu}' )\,\,(\vR\cdot \boldsymbol{\mu})&=&\Big(R_X^2 \mu'_X\mu_X + R_Y^2 \mu'_Y\mu_Y +R_Z^2 \mu'_Z\mu_Z\Big)\nn\\
&+& \Big(R_XR_Y (\mu_X\mu'_Y+\mu_Y\mu'_X) + \cdots\Big)\label{app:4}
\eea
As a result, $S_\vK(\mu',\mu)$ taken for $\vK=\bf0$ contains the following sums over $\vR$, which, for $\vR$ being a vector of the 
$({\bf X}, {\bf Y}, {\bf Z})$ cubic axes, reduce through symmetry to
\bea
\sum_{\vR\not=\bf0}\frac{R_X^2}{R^5}=\sum_{\vR\not=\bf0}\frac{R_Y^2}{R^5}=\sum_{\vR\not=\bf0}\frac{R_Z^2}{R^5}\hspace{2.5cm}\nn\\
=\frac{1}{3}\sum_{\vR\not=\bf0}\frac{R_X^2+R_Y^2+R_Z^2}{R^5}=\frac{1}{3}\sum_{\vR\not=\bf0}\frac{1}{R^3}\label{app:5}
\eea
while cross terms are equal to zero,
\be\label{app:6}
\sum_{\vR\not=\bf0}\frac{R_XR_Y}{R^5}=0
\ee

As a result, 
\bea
 S_{\vK=\bf0}(\mu',\mu)&=&\sum_{\vR\not=\bf0}\frac{1}{R^3}\left(\boldsymbol{\mu}'\cdot \boldsymbol{\mu}-\mu'_X\mu_X - \mu'_Y\mu_Y - \mu'_Z\mu_Z\right)\nn\\
 &=&0\label{app:7}
\eea

\subsubsection{Calculation for $\vK\rightarrow\bf0$ }

We first note that, since $S_{\vK=\bf0}(\mu',\mu)=0$, the $S_\vK(\mu',\mu)$ sum also reads 
\be\label{app:8}
 S_\vK(\mu',\mu)=\sum_{\vR\not=\bf0}\frac{(e^{-i\vK\cdot \vR}-1)}{R^3}\left(\delta_{\mu',\mu}-3 \frac{R_{\mu'}R_\mu}{R^2}\right)
 \ee
This shows that $S_\vK(\mu',\mu)$ for $\vK\rightarrow\bf0$ is controlled by large $R$'s. For such $R$'s, the discrete sum over cubic lattice vectors $\vR$ with lattice size $a_c$ can be replaced by integral according to
\be\label{app:9}
\sum_\vR f(\vR)\simeq \frac{1}{a_c^3}\int d^3R f(\vR)
\ee

A convenient way to calculate $S_{\vK\rightarrow \bf0}(\mu',\mu)$ is to introduce another set of orthonormal vectors $({\bf x}_\vK, {\bf y}_\vK,{\bf z}_\vK)$ with ${\bf z}_\vK=\vK/K$ and to expand $(\vR,\boldsymbol{\mu}',\boldsymbol{\mu})$ on these vectors as
\bea
\vR= R(\sin\theta \cos\varphi \,{\bf x}_\vK+ \sin\theta \sin\varphi \,{\bf y}_\vK+ \cos\theta \,{\bf z}_\vK)\label{app:10}\\
\boldsymbol{\mu}= \alpha \,{\bf x}_\vK  +\beta\,{\bf y}_\vK +\gamma\, {\bf z}_\vK  \hspace{3cm}\label{app:11}
\eea
with a similar result for $\boldsymbol{\mu}'$. This gives
\begin{widetext}
\bea
\lim_{\vK \rightarrow \bf0}S_{\vK}(\mu',\mu)\simeq \frac{1}{a_c^3}\int_0^\infty R^2 dR \int_0^\pi \sin\theta d\theta \int_0^{2\pi}d\varphi \frac{e^{-iKR \cos\theta}}{R^3} \Big[\alpha'\alpha+\beta'\beta+\gamma'\gamma \hspace{4cm}\nn\\
 -3(\alpha'\sin\theta\cos\varphi+\beta'\sin\theta \sin\varphi+\gamma'\cos\theta)(\alpha\sin\theta\cos\varphi+\beta\sin\theta \sin\varphi+\gamma\cos\theta)\Big]\label{app:12}
\eea
We first perform the integration over $\varphi$, which reduces the above equation to
\bea
\lim_{\vK \rightarrow \bf0}S_{\vK}(\mu',\mu)\simeq \frac{1}{a_c^3}\int_0^\infty \frac{ dR}{R} \int_0^\pi \sin\theta d\theta\, e^{-iKR \cos\theta}
\Big[2\pi(\alpha'\alpha+\beta'\beta+\gamma'\gamma)-3\pi (\alpha'\alpha+\beta'\beta)\sin^2\theta -6\pi \gamma'\gamma \cos^2\theta\Big]\label{app:13}
\eea
\end{widetext}
We then set $x=KR$ to get
\bea
\lim_{\vK \rightarrow \bf0}S_{\vK}(\mu',\mu)\simeq \frac{1}{a_c^3}\int_0^\infty \frac{dx}{x} \int_0^\pi \sin\theta d\theta \, e^{-ix\cos\theta}\nn\hspace{1cm}\\
\pi (- \alpha'\alpha-\beta'\beta+ 2\gamma'\gamma)(1-3\cos^2\theta)\nn\\
\simeq \frac{4\pi}{3a_c^3}(- \alpha'\alpha-\beta'\beta+ 2\gamma'\gamma)\hspace{1.8cm}\label{app:14}
\eea
where we have used
\be\label{app:15}
\int_0^\infty \frac{dx}{x}\int_{-1}^1 dt(1-3t^2)\, e^{-ixt}=\frac{4}{3}
\ee

By noting that $\gamma=\vK\cdot\boldsymbol{\mu}/K=K_\mu/K$, while $(- \alpha'\alpha-\beta'\beta+ 2\gamma'\gamma)=3\gamma'\gamma-( \alpha'\alpha+\beta'\beta+ \gamma'\gamma)$, we end with
 \be\label{app:16}
 \lim_{\vK\rightarrow \bf0}S_\vK(\mu',\mu)=-\frac{4\pi}{3a_c^3}\left(\delta_{\mu',\mu}-3 \frac{K_{\mu'}K_\mu}{K^2}\right)
 \ee
Note the sign change between the definition of $S_\vK(\mu',\mu)$ and its value for $\vK\rightarrow\bf0$. 

\subsubsection{Explicit form of the singularity}
 The above sum depends on the $\vK$ direction.
 
\noindent $\bullet$ For $\vK$ along a crystal axis, as $K_z=K$ and $K_x=K_y=0$, we find that $S_{\vK\rightarrow{\bf0}}(\mu',\mu)=0$ for $\mu'\not=\mu$, while for $\mu'=\mu$, it takes two different values: $S_{\vK\rightarrow{\bf0}}(z,z)=8\pi/3a_c^3$ and $S_{\vK\rightarrow{\bf0}}(x,x)=S_{\vK\rightarrow{\bf0}}(y,y)=-4\pi/3a_c^3$.

\noindent $\bullet$ For $\vK$ perpendicular to the $\bf z$ crystal axis, that is, $K_z=0$ and $(K_x,K_y)=K(\cos\theta,\sin\theta)$, we find $S_{\vK\rightarrow{\bf0}}(z,z)=-4\pi/3a_c^3$,  $S_{\vK\rightarrow{\bf0}}(x,x)=(4\pi/3a_c^3)(3\cos^2\theta -1)$, and $S_{\vK\rightarrow{\bf0}}(y,y)=(4\pi/3a_c^3)(3\sin^2\theta -1)$. The off-diagonal terms are $S_{\vK\rightarrow{\bf0}}(x,z)=S_{\vK\rightarrow{\bf0}}(y,z)=0$ and  $S_{\vK\rightarrow{\bf0}}(x,y)=(4\pi/a_c^3) \sin\theta\cos\theta$.

\section{Link between $P_{c,v}$ and $d_{c,v}$ \label{app:D}}

 \noindent $\bullet$ The $P_{c,v}$ term defined in Eq.~(\ref{33_1}) reads
 \bea
 \lan u_{c,\bf0}|\hat{\vp}|u_{v,\mu,\bf0}\ran&=& \int d^3 r\, \lan u_{c,\bf0}|\vr\ran\lan \vr|\hat{\vp}|u_{v,\mu,\bf0}\ran\nn\\
 &=&\int d^3 r\, \lan u_{c,\bf0}|\vr\ran\frac{\hbar}{i}\nabla \lan \vr|u_{v,\mu,\bf0}\ran \label{app:D_1}
 \eea
 which, with the help of Eqs.~(\ref{cheh_ex:3},\ref{cheh_ex:4},\ref{10'}), can be rewritten, for $\vr=\vR_\ell+\boldsymbol{\rho}$, as
 \bea
 P_{c,v}{\bf e}_\mu &=& \sum_{\vR_\ell} \int \frac{d^3 \rho}{L^3}\,  u^\ast_{c,\bf0}(\boldsymbol{\rho})\frac{\hbar}{i}\nabla  u_{v,\mu,\bf0}(\boldsymbol{\rho})\nn\\
 &=&\int \frac{d^3 \rho}{a_c^3}\,u^\ast_{c,\bf0}(\boldsymbol{\rho})\frac{\hbar}{i}\nabla  u_{v,\mu,\bf0}(\boldsymbol{\rho})\label{app:D_2}
 \eea
 
  \noindent $\bullet$ The trick is to note that in the above integral, $\nabla$ can be replaced by $(\boldsymbol{\rho}\nabla^2-\nabla^2 \boldsymbol{\rho})/2$. This leads to
  \be\label{app:D_3}
  \frac{\hbar}{im_0}P_{c,v}{\bf e}_\mu=\int \frac{d^3 \rho}{a_c^3}\,u^\ast_{c,\bf0}(\boldsymbol{\rho})\Big[\boldsymbol{\rho}, -\frac{\hbar^2}{2m_0}\nabla^2+v(\boldsymbol{\rho}) \Big]_- u_{v,\mu,\bf0}(\boldsymbol{\rho})
  \ee
 
  \noindent $\bullet$ Next, we note from Eqs.~(\ref{cheh_ex:5_1},\ref{10}) that
  \be\label{app:D_4}
  \va_{n,\bf0}\lan\vr|u_{n,\bf0}\ran= \lan\vr|h |u_{n,\bf0}\ran=\left(-\frac{\hbar^2}{2m_0}\nabla^2+v(\boldsymbol{\rho})\right)\frac{u_{n,\bf0}(\boldsymbol{\rho})}{L^{3/2}}
  \ee

 When used in Eq.~(\ref{app:D_3}), this equation readily gives
 \be
 \frac{\hbar}{im_0}P_{c,v}{\bf e}_\mu=\big( \va_{v,\bf0}- \va_{c,\bf0}\big)\int \frac{d^3 \rho}{a_c^3}\,u^\ast_{c,\bf0}(\boldsymbol{\rho})\,\boldsymbol{\rho}\,  u_{v,\mu,\bf0}(\boldsymbol{\rho})
 \ee
 where the integral is just $d_{c,v}/|e| $, while $\va_{c,\bf0}- \va_{v,\bf0}=E_{gap}$.


\begin{thebibliography}{99}
 
 \bibitem{Kittelbook} C. Kittel, \textit{Introduction to Solid State Physics}, 7th ed. (Wiley, New York, 1996).
 
 \bibitem{Bloch1929} F. Bloch, Zeitschrift f\"{u}r Physik {\bf52}, 555 (1929).

\bibitem{Merminbook} N. W. Ashcroft, and N. D. Mermin, \textit{Solid State Physics} (Holt, Rinehart and Winston, New York, 1976).

 \bibitem{Monicbook} M. Combescot and S.-Y. Shiau, \textit{Excitons and Cooper Pairs}, Oxford University Press (Oxford, 2015).

\bibitem{Haug&Koch} H. Haug and S. W. Koch, \textit{Quantum theory of the optical and electronic properties of semiconductors}, 3rd ed. (World Scientific, 1990).

\bibitem{Mahan} G. D. Mahan, \textit{Many-Particles Physics}, (Plenum Press, New York and London, 1981).


\bibitem{SYsec2021}  S.-Y. Shiau, and M. Combescot, Semiconductors {\bf55}, 1068 (2021).

\bibitem{Fetter} A. L. Fetter, and J. D. Walecka, \textit{Quantum Theory of Many Particle Systems}, (McGraw-Hill, 1971).

\bibitem{Wannier1937} G. H. Wannier, Phys. Rev. {\bf52}, 192 (1937).

\bibitem{Elliot1961} R. J. Elliot, Phys. Rev. {\bf108}, 1384 (1957); Phys. Rev. {\bf124}, 340 (1961).

\bibitem{Rashba1959} E. I. Rashba, J. Exptl. Theoret. Phys. {\bf36}, 1703 (1959) [Soviet Physics JETP {\bf36}, 1213 (1959)].

\bibitem{Moskalenko1959} S. A. Moskalenko, and K. B. Tolpygo, Zh. Eksp. Teor. Fiz. {\bf36}, 149 (1959) [Soviet Physics JETP {\bf36}, 103 (1959)].

\bibitem{Knoxbook} R. S. Knox, \textit{Theory of Excitons}, Academic Press, New York and London (1963).




\bibitem{Onodera1967} Y. Onodera, and Y. Toyozawa, J. Phys. Soc. Jpn. {\bf22}, 833 (1967).

\bibitem{Dos1968} K. Dos, A. Haug, and R. Rohner, Phys. Status Solidi (b) {\bf30}, 619 (1968).
\bibitem{Skettrup1970} T. Skettrup, and I. Balslev, Phys. Status Solidi (b) {\bf40}, 93 (1970). 
\bibitem{Pikus1971} G. E. Pikus, and G. L. Bir, Zh. Eksp. Teor. Fiz. {\bf60}, 195 (1971) [Sov. Phys. JETP {\bf33}, 108 (1971)].
\bibitem{Rohner} P. G. Rohner, Phys. Rev. B {\bf3}, 433 (1971).
\bibitem{Denisov1973} M. M. Denisov, and V. P. Makarov,  Phys. Status Solidi (b) {\bf56}, 9 (1973). 

\bibitem{Makarov1968} V. P. Makarov, Zh. Eksp. Teor. Fiz. {\bf54}, 324 (1968) [Sov. Phys. JETP {\bf27}, 173 (1968)].

\bibitem{Cho1976} K. Cho, Phys. Rev. B {\bf14}, 4463 (1976).





\bibitem{Hanamura} E. Hanamura, and H. Haug, \textit{condensation effects of excitons}, Physics Reports {\bf33}, 209 (1977).

\bibitem{Haug&Schmitt-Rink} H. Haug, and S. Schmitt-Rink, \textit{Electron theory of the optical properties of laser-excited semiconductors}, Prog. Quant. Electr. {\bf9}, 3 (1984).
 
\bibitem{Kohn1958} W. Kohn, Phys. Rev. {\bf110}, 857 (1958).


\bibitem{Abe1962} T. Abe, Y. Osaka, and A. Morita, J. Phys. Soc. Jpn. {\bf17}, 1576 (1962).

\bibitem{Sham1966} L. J. Sham, and T. M. Rice, Phys. Rev. {\bf144}, 708 (1966).


\bibitem{Cardona}P. Y. Yu, and M. Cardona, \textit{Fundamentals of Semiconductors}, 3rd ed. (Springer, Berlin, 2005).

\bibitem{DresselhausPR1955} G. Dresselhaus, A. F. Kip, and C. Kittel, Phys. Rev. {\bf 98}, 368 (1955).
\bibitem{Dresselhaus1956} G. Dresselhaus, J. Phys. Chem. Solids, {\bf1}, 14 (1956).

\bibitem{Luttinger} J. M. Luttinger, Phys. Rev. {\bf102}, 1030 (1956).

\bibitem{Baldereschi_prl} A. Baldereschi, and N. O. Lipari, Phys. Rev. Lett. {\bf25}, 373 (1970).
\bibitem{Baldereschi_prb} A. Baldereschi, and N. O. Lipari, Phys. Rev. B {\bf3}, 439 (1971).

\bibitem{SYprb2021} S.-Y. Shiau, and M. Combescot, Phys. Rev. B {\bf104}, 045203 (2021).

\bibitem{Shiauprbl2022} S.-Y. Shiau, B. Eble, and M. Combescot, Phys. Rev. B {\bf107}, L081203 (2023).


\bibitem{Ulbrich1977} R. G. Ulbrich, and C. Weisbuch, Phys. Rev. Lett. {\bf38}, 865 (1977); C. Weisbuch, and R. G. Ulbrich, \textit{Light Scattering in Solids} III, Ch. 7, p. 207 (1982, Springer). 

\bibitem{Andreani1988} L. C. Andreani, F. Bassani, and A. Quattropani, Il Nuovo Cimento {\bf10}, 1473 (1988).


\bibitem{Quattropani1986} A. Quattropani, L. C. Andreani, and  F. Bassani, Nuovo Cimento {\bf D7}, 55 (1986).

\bibitem{Bassani1986} F. Bassani, F. Ruggiero, and A. Quattropani,  Nuovo Cimento {\bf D7}, 700 (1986).

\bibitem{MoniqueEPL2022} M. Combescot, F. Dubin, and S.-Y. Shiau, Europhys. Lett. {\bf138}, 36002 (2022).

\bibitem{Sham1993} M. Z. Maialle, E. A. de Andrada e Silva, and L. J. Sham, Phys. Rev. B {\bf47}, 15776 (1993).

\bibitem{Dyakonov1986} M. I. Dyakonov and V. Y. Kachorovskii, Sov. Phys. Semicond. {\bf20}, 110 (1986).

\bibitem{Glazov2014} M. M. Glazov, T. Amand, X. Marie, D. Lagarde, L. Bouet, and B. Urbaszek, Phys. Rev. B {\bf89}, 201302(R) (2014).

\bibitem{Qiu2015}D. Y. Qiu, T. Cao, and S. G. Louie, Phys. Rev. Lett. {\bf115}, 176801 (2015).
 
\bibitem{Echeverry2016}J. P. Echeverry, B. Urbaszek, T. Amand, X. Marie, and I. C. Gerber, Phys. Rev. B {\bf93}, 121107(R) (2016).

\bibitem{Pengkeprb2022}P. Li, C. Robert, D. V. Tuan, L. Ren, M. Yang, X. Marie, and H. Dery, Phys. Rev. B {\bf106}, 085414 (2022).

\end{thebibliography}
\end{document}